\documentclass[10pt]{iopart}
\pdfoutput=1
\usepackage{iopams}
\usepackage{graphicx}
\usepackage{epstopdf}
\usepackage{wrapfig}
\usepackage{bibentry}
\usepackage{multicol}
\usepackage{todo}
\usepackage{xcolor}
\usepackage{enumerate}
\usepackage[english]{babel}
\usepackage{float}
\usepackage{amsfonts}
\usepackage{wrapfig}
\usepackage{setspace}

\usepackage{scalerel}
\usepackage{stackengine}
\usepackage{wasysym}

\newcommand\reallywidetilde[1]{\ThisStyle{%
		\setbox0=\hbox{$\SavedStyle#1$}%
		\stackengine{-.1\LMpt}{$\SavedStyle#1$}{%
			\stretchto{\scaleto{\SavedStyle\mkern.2mu\AC}{.5\wd0}}{.3\ht0}%
		}{O}{c}{F}{T}{S}%
}}

\begin{document}
\title[New insights on divertor flows, drifts, and fluctuations from in situ, 2D probe measurement in TCV]
{New insights on divertor parallel flows, ExB drifts, and fluctuations from in situ, two-dimensional probe measurement in the Tokamak à Configuration Variable}

\author{H. De Oliveira$^1$, C. Theiler$^1$, O. F\'evrier$^1$, H. Reimerdes$^1$, B. P. Duval$^1$, C. Tsui$^2$, S. Gorno$^1$, D. S. Oliveira$^1$, A. Perek$^3$, the TCV Team$^a$}
\address{$^1$\'Ecole Polytechnique F\'ed\'erale de Lausanne (EPFL), Swiss Plasma Center (SPC), CH-1015 Lausanne, Switzerland.}
\address{$^2$Center for Energy Research (CER), University of California San Diego (UCSD), La Jolla, California 92093-0417, USA}
\address{$^3$Dutch Institute for Fundamental Energy Research, De Zaale 20, 5612 AJ Eindhoven, Netherlands.}
\address{$^a$See the author list of H. Reimerdes et al 2022 \emph{Nucl. Fusion} \textbf{62} 042018.}
\ead{christian.theiler@epfl.ch}
\date{\today}
\begin{abstract}
In-situ, two-dimensional (2D) Langmuir probe measurements across a large part of the TCV divertor are reported in L-mode discharges with and without divertor baffles. This provides detailed insights into time averaged profiles, particle fluxes, and fluctuation behavior in different divertor regimes. The presence of the baffles is shown to substantially increase the divertor neutral pressure for a given upstream density and to facilitate the access to detachment, an effect that increases with plasma current. The detailed, 2D probe measurements allow for a divertor particle balance, including ion flux contributions from parallel flows and $E\times B$ drifts. The poloidal flux contribution from the latter is often comparable or even larger than the former, such that the divertor parallel flow direction reverses in some conditions, pointing away from the target. In most conditions, the integrated particle flux at the outer target can be predominantly ascribed to ionization along the outer divertor leg, consistent with a closed-box approximation of the divertor. The exception is a strongly detached divertor, achieved here only with baffles, where the total poloidal ion flux even decreases towards the outer target, indicative of significant plasma recombination. The most striking observation from relative density fluctuation measurements along the outer divertor leg is the transition from poloidally uniform fluctuation levels in attached conditions to fluctuations strongly peaking near the X-point when approaching detachment.
\end{abstract}
\submitto{\NF}

\pacs{}
\maketitle

\ioptwocol

\section{Introduction}\label{sec:introdution}
Particle and heat exhaust in next generation tokamaks such as ITER and DEMO remains an outstanding issue that must be addressed in order to maintain plasma facing components in working condition whilst maintaining acceptable core performance. The transport of energy and particles within the boundary of tokamaks is governed by a complex interplay of cross-field turbulent transport and steady-state drifts, strong transport parallel to the magnetic field together with sources and sinks. With this complexity, extrapolation towards higher power tokamaks is challenging, requiring detailed experimental insights and comparison with theory on today's experiments. TCV, a tokamak \cite{Reimerdes_2022} at EPFL, is a graphite wall machine with exceptional magnetic poloidal shaping capabilities. It can access a wide range of configurations, that may be used to shed light upon the complex boundary dynamics and assist in model validation.

\begin{figure}[h!]
	\centering
	\includegraphics[width=\linewidth]{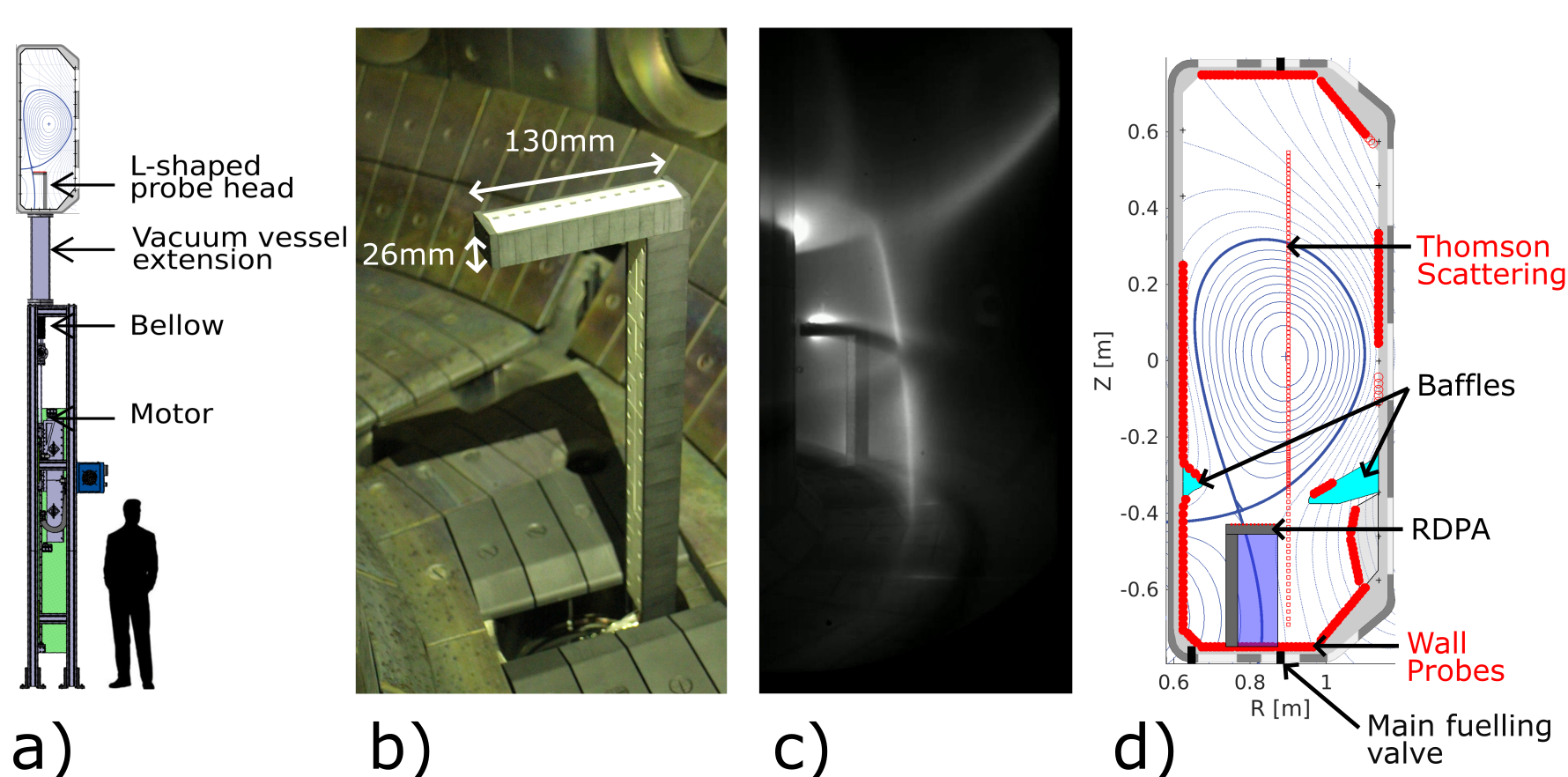}
	\caption{\label{fig:diagnostic_coverage} {a)} Poloidal view of the RDPA diagnostic structure below TCV, {b)} RDPA in vessel picture taken during machine opening, {c)} $D4\rightarrow 2$ light snapshot recorded with the MANTIS camera system \cite{Perek2019} during an experiment and {d)}  Illustration of the magnetic geometry used in this work together with the baffles, gas valves and the spatial coverage of the most relevant diagnostics used in this study: Thomson Scattering (TS), RDPA and wall probes. The blue shaded rectangle represents the region accessed by RDPA during an up/down reciprocation.}
\end{figure}

In this work, which is an extension of the work illustrated in Chapters 4 and 6 of Ref. \cite{DeOliveiraThesis}, we report on two-dimensional (2D) Langmuir probe measurements across a large part of TCV's divertor region, providing a unique experimental approach to the measurement of divertor particle fluxes..  The measurements are enabled by a new, fast Reciprocating Divertor Langmuir Probe Array (RDPA) \cite{DeOliveira2021}. Specifically, in the divertor geometry studied in this work, RDPA provides 2D profiles along the outer divertor leg, from the combination of a radial array of probes and an extended vertical sweep, Figure \ref{fig:diagnostic_coverage}. The quantities obtained include plasma density, electron temperature, potential, parallel ion Mach number, and density fluctuation levels. They are used to perform a particle balance analysis in the TCV divertor for different divertor regimes with and without TCV's divertor baffles $\cite{Reimerdes2017,Fasoli2020}$. The particle balance study reveals that $E \times B$ drifts often play an important role, as previously observed in DIII-D \cite{Boedo2000,Boedo1999,Jarvinen2019}, and that a closed-box approximation \cite{Krasheninnikov1987,Krasheninnikov2017} is a relatively good assumption for the TCV divertor, except for when the plasma is strongly detached. Another interesting observation is a reversal of the parallel flow in the downward $E\times B$ region in certain conditions. Regarding divertor density fluctuations, a key observation is the appearance of a strong poloidal gradient of the turbulence level at high collisionality, where the fluctuation near the target become much weaker than near the X-point.

The paper is organized as follows. A description of the experimental setup is presented in Section $\ref{sec:experimental_setup}$, followed by the analysis performed to deduce the most relevant quantities from the RDPA in Section $\ref{sec:flows_calculation}$. The main experimental results are presented in Section $\ref{sec:particle_balance}$ and Section $\ref{sec:fluctuation_results}$. Finally, a discussion and summary of the paper is presented in Section $\ref{sec:conclusion}$.

\section{Experimental setup}\label{sec:experimental_setup}
\begin{figure*}
\centering
\includegraphics[width=\linewidth]{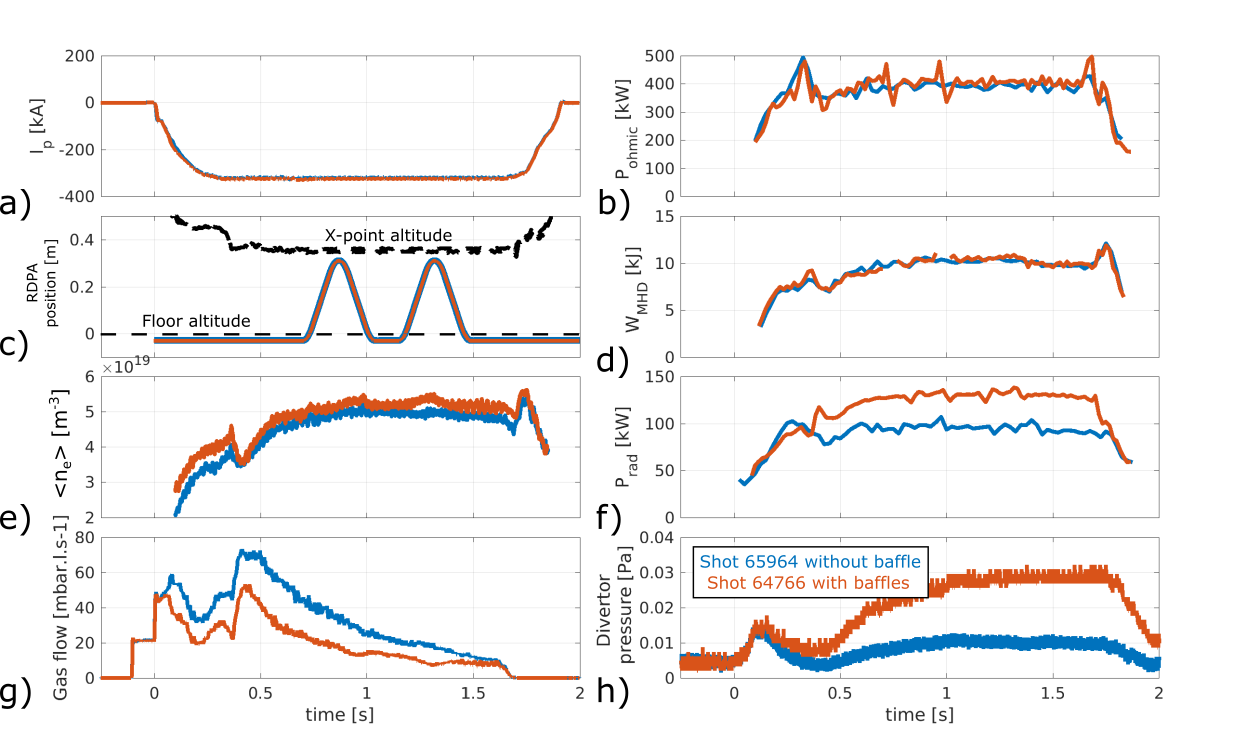}
\caption{\label{fig:figure_time_traces_RDPA_effect}  Time traces of relevant plasma properties from the low density reference $\#65964$ without baffles and the low density reference $\#64766$ with baffles: {a)} plasma current, {b)} ohmic power (no external heating source has been used in this study), {c)} vertical position of the RDPA probe and X-point height, {d)} stored energy computed from magnetic measurements, {e)} line averaged density from interferometry measurements, {f)} total radiated power from bolometry measurements, {g)} deuterium gas flow and {h)} divertor neutral pressure from baratron measurements.}
\end{figure*}

The present divertor particle balance studies are performed in Lower Single Null (LSN) Ohmic L-mode configurations with a toroidal magnetic field $B_t \approx 1.4~\mathrm{T}$ and constant plasma current $I_p \approx 320~\mathrm{kA}$, Figure \ref{fig:diagnostic_coverage}{d)}. The toroidal field is in the so-called 'reversed' direction, corresponding to an upward ion $\vec{B} \times {\nabla \vec{B}}$ drift that is unfavorable to H-mode access. $I_p=320~\mathrm{kA}$ is chosen, resulting in a $q_{95}$ of  $\approx 2.4$. This choice results in magnetic field lines with relatively large pitch angle, limiting the total fraction of field lines that are intercepted by the RDPA horizontal probe arm in the divertor outer leg to less than 15\%. A high plasma current also permits a higher line averaged density. In these experiments, the line averaged core density $\left<n_e\right>$was varied from $5\times 10^{19}~\mathrm{m^{-3}}$ up to $13\times 10^{19}~\mathrm{m^{-3}}$ to access both attached and detached divertor conditions. 

Edge physics experiments are often conducted with density ramps in TCV $\cite{Theiler2017}$ to study the plasma evolution from attached to detached within the same discharge. Since RDPA vertical plunges last typically $\approx 0.35~\mathrm{s}$ (limited by the motor power), the measurements during a density ramp would not provide 2D divertor profiles with consistent conditions. Therefore, multiple discharges were performed with constant core densities. The RDPA was found to have a negligible effect on the relevant main plasma properties such as radiated power, stored energy, line averaged density, Ohmic power, gas fueling and divertor pressure. These quantities remain constant across the reciprocation period, as shown in Figure $\ref{fig:figure_time_traces_RDPA_effect}$. RDPA produces, however, a characteristic shadow downstream, with a plasma density lower than in the surrounding and data from the floor LPs influenced by this shadow were discarded.

\begin{figure*}
\centering
\includegraphics[width=\linewidth]{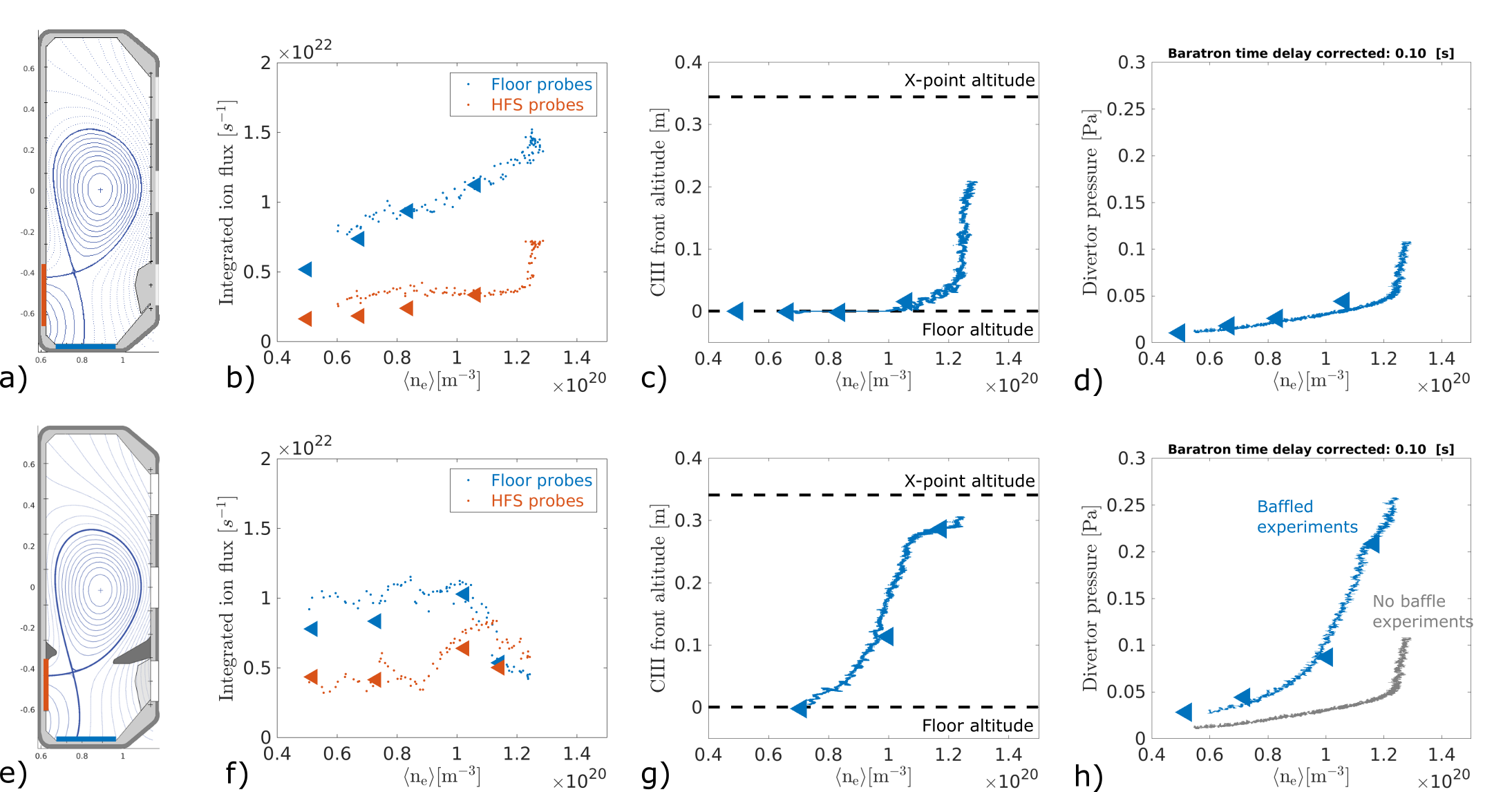}
\caption{\label{fig:density_comparison} {a)} and {e)} Magnetic equilibrium reconstruction and relevant Langmuir probe coverage highlighted on the walls, {b)} and {f)} Integrated outer target ion flux as a function of $\langle n_e \rangle$, {c)} and {g)} CIII front altitude obtained from the multi spectral imaging MANTIS system \cite{Perek2019} and {d)} and {e)} divertor neutral pressure obtained from the baratron gauge. Shots without baffles are shown in {a)}, {b)}, {c)} and {d)}: constant density experiments \#65964, \#66210, \#66220 and \#66222 represented with triangles and the density ramp \#66208 represented with scattered points. Shots with baffles are shown in {e)}, {f)}, {g)} and {h)}: constant density experiments \#64766, \#63963, \#64965 and \#64962 represented with triangles and the density ramp \#64900 with scattered points. CIII data for the shot \#64766 could not be acquired.}
\end{figure*}

The total integrated ion flux density at the outer target, the CIII emission front position (a proxy for low electron temperature conditions) along the outer leg \cite{Theiler2017} and the divertor neutral pressure, all given as a function of the line averaged density, are found similar with density ramps and with constant density, as shown in Figure $\ref{fig:density_comparison}$. This indicates that the divertor conditions, specifically including neutral dynamics and impurity levels, can be considered close to equilibrium for any time in density ramp experiments. 

Recently, the heat and particle exhaust characteristics of the TCV tokamak were modified\cite{Reimerdes2021,Fevrier2021} by an in-vessel structures of solid graphite baffles that form a divertor chamber of increased closure, to better decouple divertor and main chamber regions. In this work, experiments were performed with an open divertor (non-baffled), see Figure $\ref{fig:density_comparison}${a)}, {b)}, {c)} and {d)}, and in the presence of the first version of the baffles, see Figure $\ref{fig:density_comparison}${e)}, {f)}, {g)} and {h)}. The integrated ion flux does not roll over during the density ramps in the absence of the baffles for these shots, see Figure $\ref{fig:density_comparison}$ {b)}, in contrast to the baffled shots, with a roll over at $\langle n_e \rangle \lesssim 10\times 10^{19}~\mathrm{m^{-3}}$, see Figure $\ref{fig:density_comparison}${f)}. Likewise, with the baffles an earlier movement of the CIII front towards the X-point and a substantially higher neutral divertor pressure (up to a factor $5$ higher) is obtained for the same core conditions. These effects of the baffles are similar, but even stronger than previous L-mode results obtained at $I_p=250~\mathrm{kA}$ \cite{Reimerdes2021,Fevrier2021}.

Note that the gas flow required for fueling was lower for the baffled low density discharge in Figure $\ref{fig:figure_time_traces_RDPA_effect}${g)} than for the non-baffled one. This isnot a general observation with the gas flow in other experiments being often higher with baffles installed. The result in Figure $\ref{fig:figure_time_traces_RDPA_effect}${g)} can be explained by the dominant plasma fueling source coming from recycled neutrals, such that small changes in the recycling coefficient, a property of the wall surface condition, can substantially influence the required gas puffing rate.

\section{Procedure for the interpretation of RDPA measurements}\label{sec:flows_calculation}
In this section, we describe how key quantities, such as the parallel ion Mach number, electron density, ion flux along the magnetic field and the $E\times B$ fluxes, are deduced from time averaged RPDA  measurements. To illustrate these steps, example results from RDPA in an attached L-mode plasma without baffles, at a line-averaged density of $\langle n_e \rangle \approx 6.75\times 10^{19}~\mathrm{m^{-3}}$, and with same geometry and experimental parameters as for the discharges in Figure \ref{fig:density_comparison} (unfavorable ion $\vec{B} \times \nabla \vec{B}$ drift, $I_p\approx 320~\mathrm{kA}$) are presented in Figures \ref{fig:reference_flows} to \ref{fig:reference_ExB_compare}.

\begin{figure}
\centering
\includegraphics[width=\linewidth]{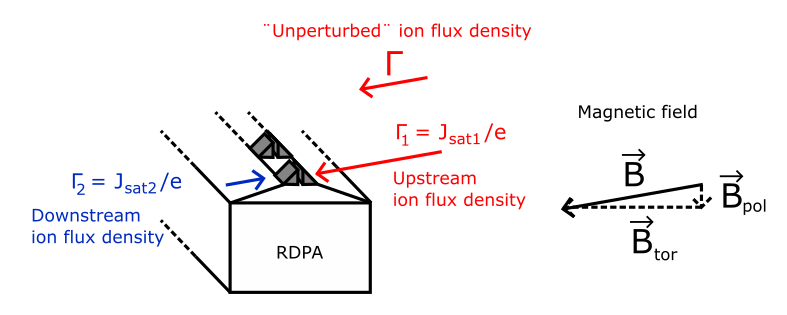}
\caption{\label{fig:scheme_Mach_probes} Sketch of the RDPA probe arm with embedded probes and their position relative to the magnetic field: upstream or downstream.}
\end{figure}

The horizontal probe arm of the RDPA, shown in Figure \ref{fig:diagnostic_coverage}{b)}, is equipped \cite{DeOliveira2021} with $12$ Mach probes, radially spaced by $10~\mathrm{mm}$,  Figure $\ref{fig:scheme_Mach_probes}$. These probe tips can be operated at constant bias, in voltage sweep, or in floating potential mode, similarly to the TCV wall-embedded probes \cite{Fevrier2018,DeOliveira2019}. Data is acquired at $2~\mathrm{MHz}$. When operated in swept mode, and for each probe tip, the ion saturation current $J_{sat}$, the electron temperature $T_e$, and the floating potential $V_{fl}$ are obtained from a $4$-parameter fit that takes sheath expansion into account $\cite{Fevrier2018}$. This is generally performed upon data acquired during a $10~\mathrm{ms}$ period, equivalent to $10$ voltage sweep periods for the typical sweep frequency of $1000~\mathrm{Hz}$ and improves the data quality as compared to using data from individual sweeps.

\subsection{Parallel flow measurements.}\label{sec:parallel_flows_calculation}

\begin{figure*}
\centering
\includegraphics[width=\linewidth]{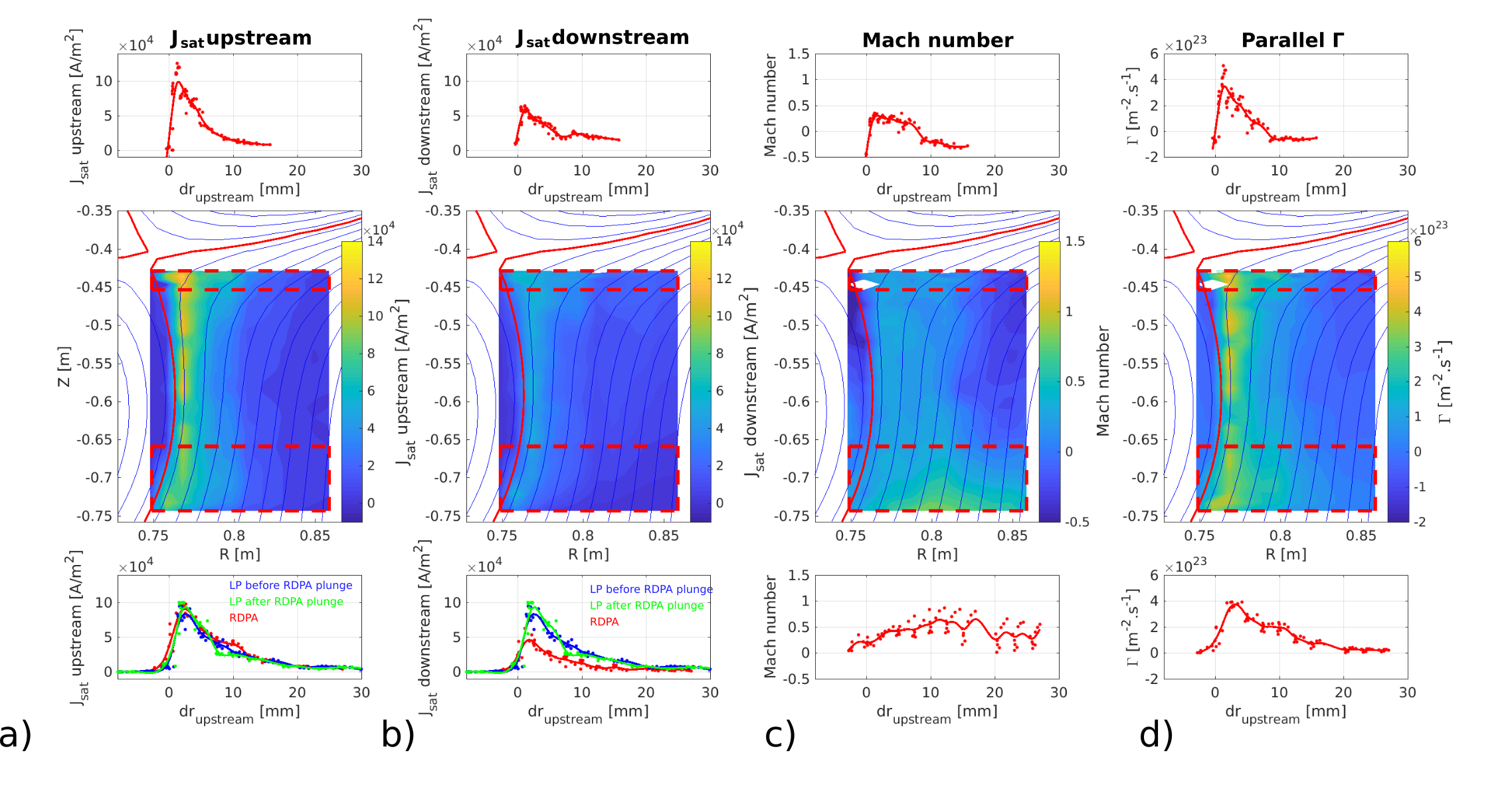}
\caption{\label{fig:reference_flows} RDPA measurements of {a)} Upstream ion saturation current density, {b)} Downstream ion saturation current density, {c)} Mach number and \textbf{d)} Parallel particle flux density. Each figure is divided into three regions: {top)} 1D profile obtained with RDPA near the top of the plunge (in the region within the dashed red rectangle near the X-point in the middle plot), {middle)} 2D contour plot from RDPA and {bottom)} 1D profile near the bottom of the plunge (in the region within the dashed red rectangle near the target in the middle plot). The blue and green data in the bottom panels of {a)} and {b)} are obtained from the floor LPs. Discharge \#63029.}
\end{figure*}

The parallel ion Mach number is deduced with RDPA from the ratio of upstream to downstream $J_{sat}$ $\cite{Hutchinson2005}$:
\begin{equation} 
M =\ln(\frac{J_{sat1}}{J_{sat2}})/2.2.
\end{equation}
Here, $J_{sat1}$ and $J_{sat2}$ are obtained from swept probes or, for a higher time resolution, with the probes operated in ion-saturation current mode.

The electron density $n_e$, in the presence of flows, is calculated from a viscous plasma model based on diffusive perpendicular transport $\cite{Hutchinson2005}$: 
\begin{equation} \label{eq:density_Hutchinson}
n_e=\frac{J_{sat1}/e}{c_s \exp(-1+1.1 M)},
\end{equation}
where $c_s$ is the plasma sound speed, usually calculated as $\sqrt{(T_e+T_i)/m_i}$ $\cite{Fevrier2018}$ with the assumption $T_e=T_i$. Unlike the sign convention given in \cite{Hutchinson2005}, here $M$ is defined as positive for a particle flux directed towards the upstream probe tip and away from the downstream probe tip. A positive Mach number, therefore, corresponds to a plasma flow directed towards the target. Equation (\ref{eq:density_Hutchinson}) shows reasonable agreement with expressions typically used for density measurements with LPs in the bulk plasma\cite{Stangeby2000} and for $M=0$, as well as for measurements at the walls (from wall-embedded probes) assuming $M=1$. For $M=0$, Equation  (\ref{eq:density_Hutchinson}) becomes:
\begin{equation}\label{eq:density_01}
n_e=\frac{J_{sat1}/e}{c_s exp(-1)}\approx\frac{2.7J_{sat1}/e}{c_s},
\end{equation}
that agrees within $35\%$ with the usual expression\cite{Stangeby2000}:
\begin{equation}
n_e=\frac{2J_{sat1}/e}{c_s}. 
\end{equation}
For $M=1$, Equation (\ref{eq:density_Hutchinson}) becomes:
\begin{equation}\label{eq:density_03}
n_e=\frac{J_{sat1}/e}{c_s \exp(0.1)}\approx 0.9\frac{J_{sat1}/e}{c_s},
\end{equation}
which is close to the commonly used formula for the sheath edge density:
\begin{equation}
n_{se}=\frac{J_{sat1}/e}{c_s}. 
\end{equation}
The density $n_e$ in Equations (\ref{eq:density_01})-(\ref{eq:density_03}) is intended to represent the plasma density of the unperturbed plasma (in the absence of the probe) and differs from the sheath edge densities $n_{se1}$ and $n_{se2}$ on the upstream and downstream probe tips. Another model, derived with drift based perpendicular transport, gives a similar formula for the density \cite{Hutchinson2008}. The contributions from perpendicular drifts (represented by the Mach number $M_\perp$ perpendicular to the magnetic field) to the calculated plasma density were calculated with this drift based model \cite{Hutchinson2008}:
\begin{equation} \label{eq:ExB_Hutchinson}
n_e=\frac{J_{sat1}/e}{c_s \exp(-1-(M-M_{\perp}cot(\theta)))},
\end{equation}
where $\theta$ is the angle (in the plane of magnetic field and drift velocity) of the object surface to the magnetic field and $M$ is defined here with the sign convention given in \cite{Hutchinson2005}. As stated in \cite{Hutchinson2008}: \textit{"If a facet lies in a concave region of the object [...], then it does not possess its own plasma region. Instead, the solution(s) of the earlier region(s) applies right up to the respective fractions of that facet."} and, therefore, the relevant angle $\theta$ for the RDPA probe tip is not the angle of the probe tip surface, because of the concavity of the RDPA cross-sectional geometry for a downward $E\times B$ velocity. Instead, the angle between the boron nitride thermal shield surface and the magnetic field direction \cite{DeOliveira2021}, $\theta \approx 15^\circ$, is appropriate for Equation (\ref{eq:ExB_Hutchinson}). The term $M_{\perp} cot(\theta)$ contributes a noticeably to the density calculation in the case of large $E\times B$ velocities, such as for the experimental results shown in Figure \ref{fig:reference_Te_Vfl} and \ref{fig:reference_ExB} ($T_e\approx 27~\mathrm{eV}$, $v_{E\times B}\approx 3000~\mathrm{ms^{-1}}$):
\begin{equation}
M_{\perp} cot(\theta) = \frac{v_{E\times B}}{c_s} cot(\theta) \leq \frac{3.7 \times 3000~\mathrm{ms^{-1}}}{51000~\mathrm{ms^{-1}}} \approx 0.22
\end{equation}
This density correction, due to the perpendicular velocity, has not been included in the present work and may lead to an error of up to $20\%$ in the density calculation for the worst case (region of strong downward $E\times B$ velocity).

Finally, the parallel ion flux density is calculated as follows: 
\begin{equation} \label{eq:flow_hutchinson}
\Gamma_\parallel =v_{\parallel}\cdot n_e = M\cdot c_s \cdot n_e=\frac{MJ_{sat1}/e}{\exp(-1+1.1 M)}.
\end{equation}
Note that the flow measurement is independent of the temperature measurement. This is an advantage in the case of detached plasmas, where the electron temperature can be overestimated for Langmuir probe measurements \cite{Fevrier2018}. The upstream ion saturation current collected by RDPA is usually similar to the ion saturation current collected by wall probes, see the bottom panel of Figure $\ref{fig:reference_flows}${a)}. The downstream current is, however, usually a factor $2$ lower, see Figure $\ref{fig:reference_flows}${b)}, resulting in a Mach number, $M\approx ln(2)2)/2.2 \approx 0.3$, as shown in Figure $\ref{fig:reference_flows}${c)}, that increases towards the target. An example of parallel ion flow measurements based on Equation (\ref{eq:flow_hutchinson}) is shown in Figure $\ref{fig:reference_flows}${d)}.

\subsection{$E\times B$ flow measurements.}\label{sec:EtimesB_flows}

\begin{figure*}
\centering
\includegraphics[width=\linewidth]{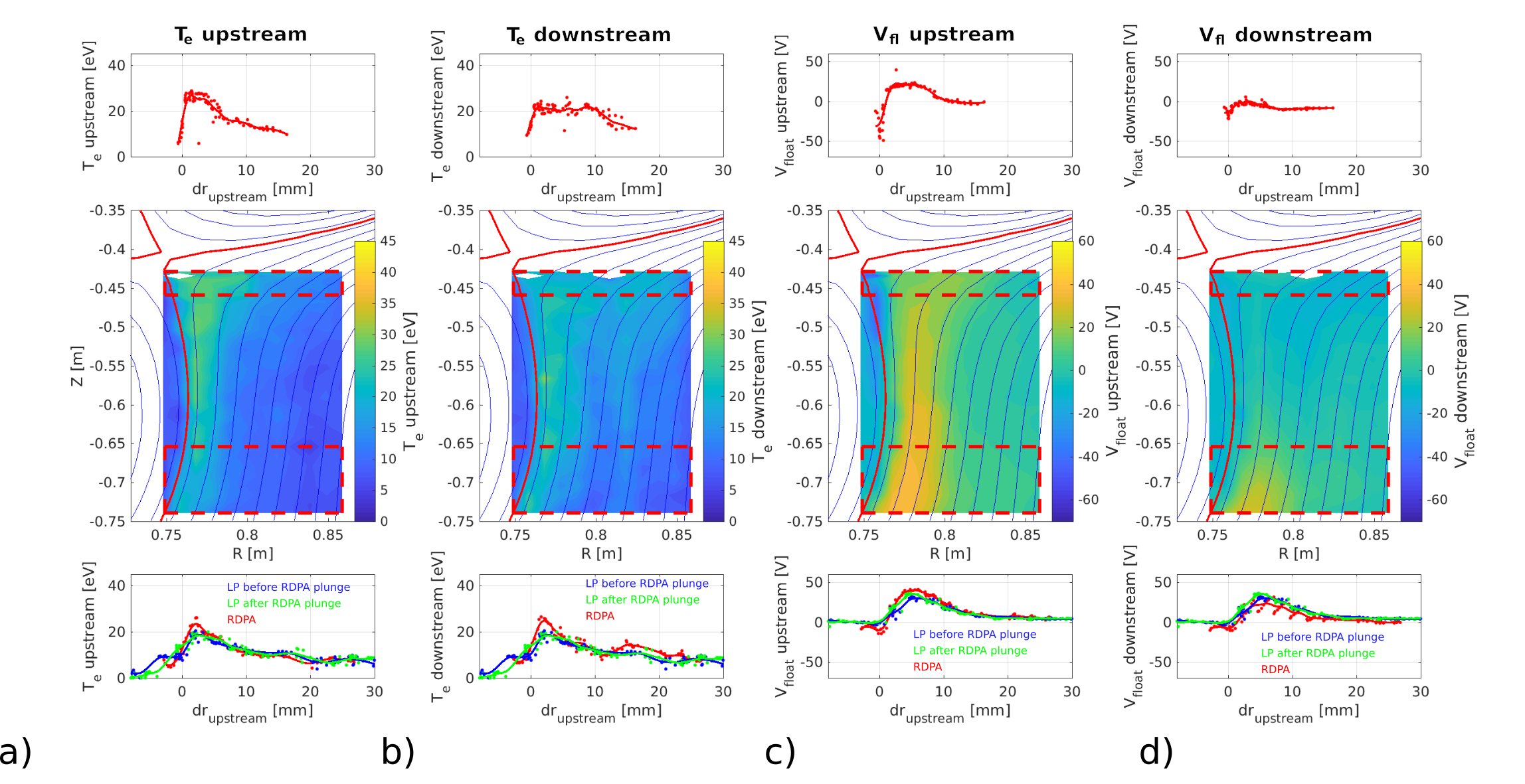}
\caption{\label{fig:reference_Te_Vfl}  Figure equivalent to Figure \ref{fig:reference_flows}, showing RDPA measurements of {a)} Upstream electron temperature, {b)} Downstream electron temperature, {c)} Upstream floating potential and {d)} Downstream floating potential. The blue and green data in the bottom panels of {a)} and {b)} are obtained from the floor LPs. Discharge \#63029.}
\end{figure*}
\begin{figure*}
\centering
\includegraphics[width=\linewidth]{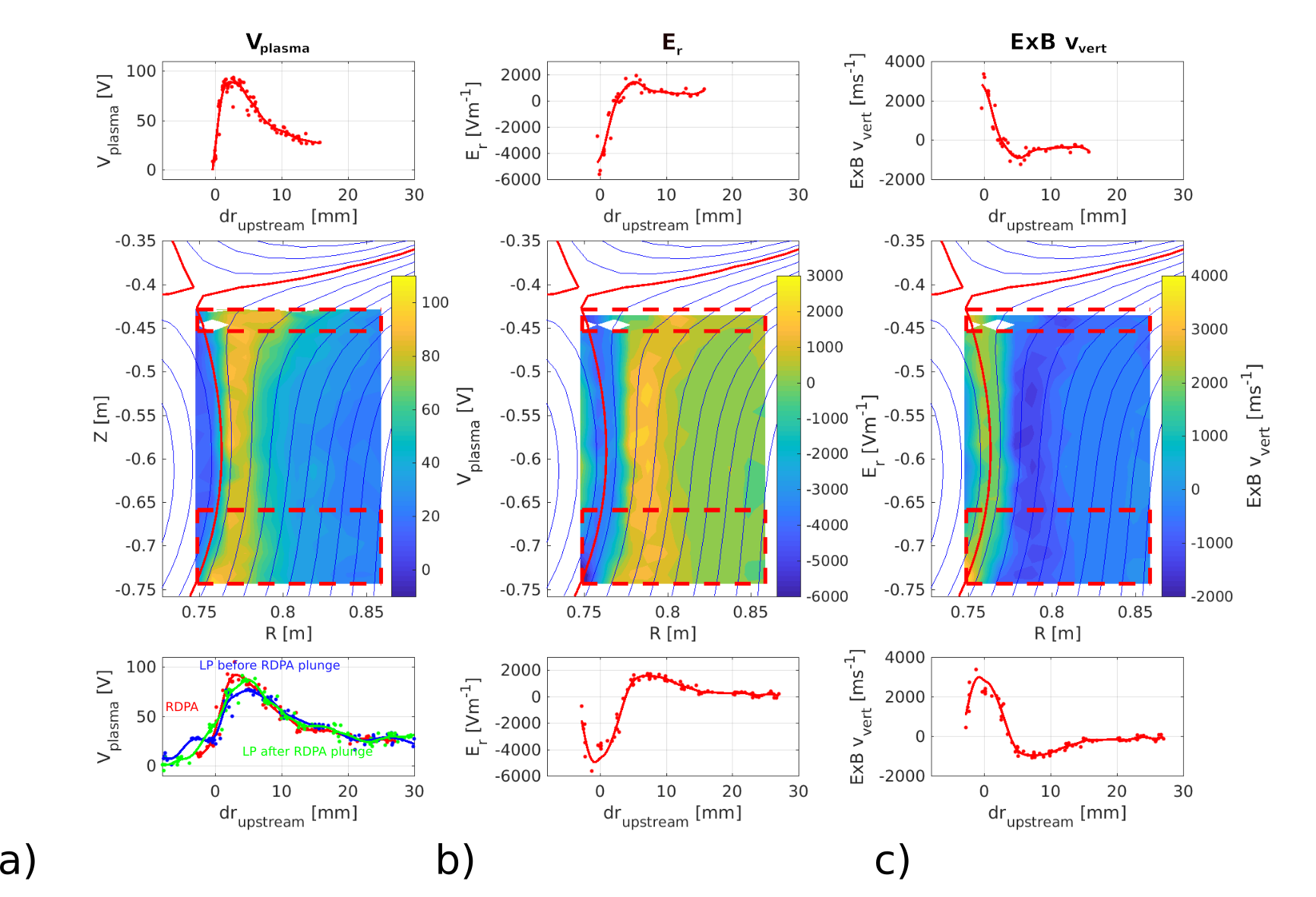}
\caption{\label{fig:reference_ExB} Figure equivalent to Figure \ref{fig:reference_flows}, showing RDPA measurements of {a)} Plasma potential, {b)} Radial electrical field and {c)} Vertical $E\times B$ velocity. Discharge \#63029.}
\end{figure*}

$E\times B$ particle fluxes are calculated as the product between the local plasma density and the $E \times B$ drift velocity:
\begin{equation}
\vec{\Gamma}_{E\times B} = n_e \cdot \vec{v}_{E\times B}=\frac{J_{sat1}/e}{c_s \exp(-1+1.1 M)} \frac{\vec{E}\times\vec{B}}{B^2}
\end{equation}
The drift velocity is calculated from the plasma potential measurement obtained from sheath theory:
\begin{equation}
V_{pl} = (V_{fl,downstream} + V_{fl,upstream})/2 + 3 T_{e,upstream},
\end{equation}
where $T_{e,upstream}$ is expressed in $[eV]$ and the constant $3$ corresponds to a deuterium plasma (assuming $T_e\approx T_i$ and no secondary electron emission \cite{Stangeby2000}). Example measurements of these quantities are presented in Figures \ref{fig:reference_Te_Vfl} and \ref{fig:reference_ExB}. In the absence of a valid theory explaining occasionally observed discrepancies between upstream and downstream floating potential measurements, apparent in Figure \ref{fig:reference_Te_Vfl}{c)} and {d)}, an average of the upstream and downstream values is taken. This choice is motivated by a quantitative agreement with the $V_{pl}$ target profiles, see Figure $\ref{fig:reference_ExB}${a)}, although it does not address the discrepancies in upstream and downstream $V_{fl}$ seen near the top of the plunge, where a comparison with wall LPs is not meaningful.

For electron temperatures, the difference between upstream and downstream values is usually less pronounced than for the $V_{fl}$ profiles, see Figure $\ref{fig:reference_Te_Vfl}${a)} and {b)}. The upstream electron temperature value was chosen in the calculations.

The electric field in the toroidal direction can be neglected as the loop voltage, in steady state conditions, is small. For the case of a predominantly toroidal field ($B_z$, $B_r\ll B_{\phi}$), the $E\times B$ drift velocity can then be written as:
\begin{equation}
\vec{v}_{E\times B} \approx \frac{-E_zB_{\phi}}{B^2}\vec{e}_r + \frac{E_rB_{\phi}}{B^2}\vec{e}_z.
\end{equation}
The second term dominates in the case considered here, with a near-vertical divertor leg and with stronger $V_{pl}$ variations across, than along, the divertor leg. In Figure \ref{fig:reference_ExB}, example 2D measurements of $V_{pl}$, $E_r$, and $\vec{v}_{E\times B}$ projected along the vertical direction are shown. The vertical $E\times B$ velocity in Figure \ref{fig:reference_ExB} is taken positive when directed towards the target corresponding to a projection along the vector $-\vec{e_z}$. This graphical choice provides the same sign convention as the parallel ion flux in Figure \ref{fig:reference_flows}{d)} (positive towards the target).


\begin{figure*}
\centering
\includegraphics[width=\linewidth]{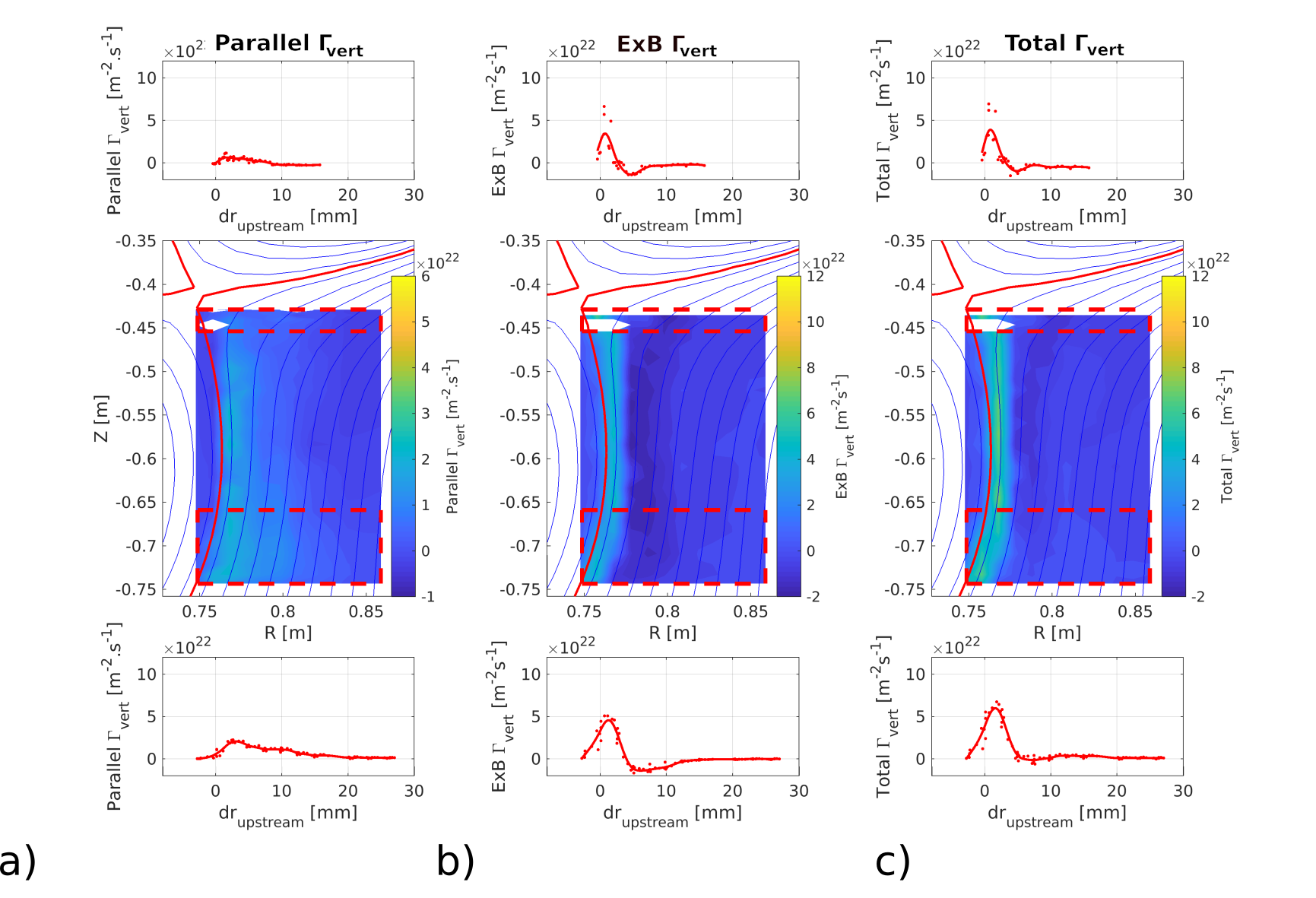}
\caption{\label{fig:reference_ExB_compare} Figure equivalent to Figure \ref{fig:reference_flows}, showing RDPA measurements of {a)} Vertical particle flux density from the parallel Mach measurement, {b)} Vertical $E\times B$ particle flux density and {c)} Total vertical particle flux density.  Discharge \#63029.}
\end{figure*}

A clear result from these measurements is that $E\times B$ flows, although much weaker than parallel flows in the divertor, bring a substantial contribution to the total poloidal ion flux due to the shallow pitch angle of the field lines (field lines are near toroidal). This poloidal contribution can be larger than that from the parallel flows, in the case of attached discharges, Figure $\ref{fig:reference_ExB_compare}$, where the high electron temperatures lead to steep plasma potential profiles. A similar observation of strong $E\times B$ flows was observed experimentally in DIII-D \cite{Boedo2000,Boedo1999,Jarvinen2019,Jarvinen2018}. 

\subsection{Density fluctuations}\label{sec:density_fluctuations_calculation}
Performing electron temperature, or even ion temperature, measurements at the plasma turbulence timescale is not possible with the conventional single Langmuir probe electronics used in this study. Density fluctuation measurements are more accessible, as $J_{sat}\propto c_s n_e$ fluctuations are generally dominated by density fluctuations \cite{Nold2012,Mahdizadeh2005}. This can be partially explained by the relatively weak dependence of $c_s$ on the electron temperature (square root). 

Density fluctuation levels are deduced here from the standard deviation of the $I_{sat}$ time windows. The typical duration of a time window is chosen to be approximately $1~\mathrm{ms}$ (equivalent to $2000$ acquired time points at a $2~\mathrm{MHz}$ frequency). The time windows can be recorded while operating the probe in $I_{sat}$ mode or in voltage sweeping mode, retaining only the intervals when the probe voltage is sufficiently negative for the probe to record $I_{sat}$. In some regions, in particular in the private flux region where signal levels are very low, the $\widetilde{J_{sat}}$ and $\widetilde{n_e}$ signals (the tilde symbol stands for their standard deviation) are dominated by instrumental noise. To correct for this, the standard deviation of the noise is recorded before the plasma discharge and then subtracted assuming that the signal and the noise are independent random variables. The variance of the actual signal is thus deduced as:
\begin{equation}
\sigma_{signal}=\sqrt{\sigma_{measurement}^2-\sigma_{noise}^2}
\end{equation}
The density fluctuations are then derived from Equation (\ref{eq:density_Hutchinson}):
\begin{equation}
\widetilde{n_e} = \reallywidetilde{\left(\frac{J_{sat1}/e}{c_s \exp(-1+1.1 M)}\right)},
\end{equation}
which we approximate as:
\begin{equation}
\widetilde{n_e}\approx\frac{1}{\langle c_s \rangle}\reallywidetilde{\left(\frac{J_{sat1}/e}{\exp(-1+1.1 M)}\right)},
\end{equation}
where $\langle c_s \rangle$ is the time-averaged, local sound speed, obtained from processing the time averaged data from swept probes. The normalized fluctuation level is approximated as:
\begin{equation}
\frac{\widetilde{n_e}}{\langle n_e \rangle} = \reallywidetilde{\left(\frac{J_{sat1}/e}{\exp(-1+1.1 M)}\right)} / \frac{\langle J_{sat1}\rangle/e}{\exp(-1+1.1 \langle M \rangle)}.
\end{equation}

\section{Particle balance results in the divertor}\label{sec:particle_balance}
In this section, we apply the RDPA data analysis presented in section \ref{sec:flows_calculation} to shed light on the divertor particle balance as a function of core density, comparing discharges without and with baffles 
\subsection{The influence of core density on the divertor particle balance - no baffles}\label{sec:particle_balance_no_baffle}
\begin{figure}
\centering
\includegraphics[width=\linewidth]{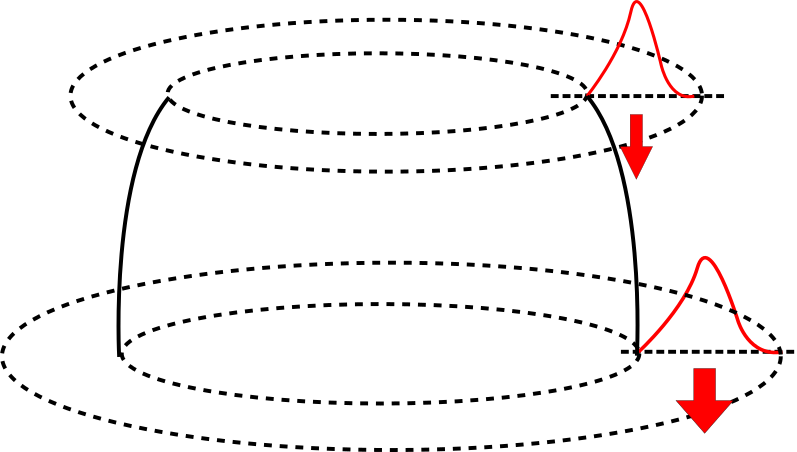}
\caption{\label{fig:integration_drawing} Sketch of the evaluation of the total ion flux to the outer target obtained by integration in the radial and toroidal directions. The red curve represents the vertical ion flux due to parallel and $E \times B$ flows.}
\end{figure}

\begin{figure*}
\centering
\includegraphics[width=\linewidth]{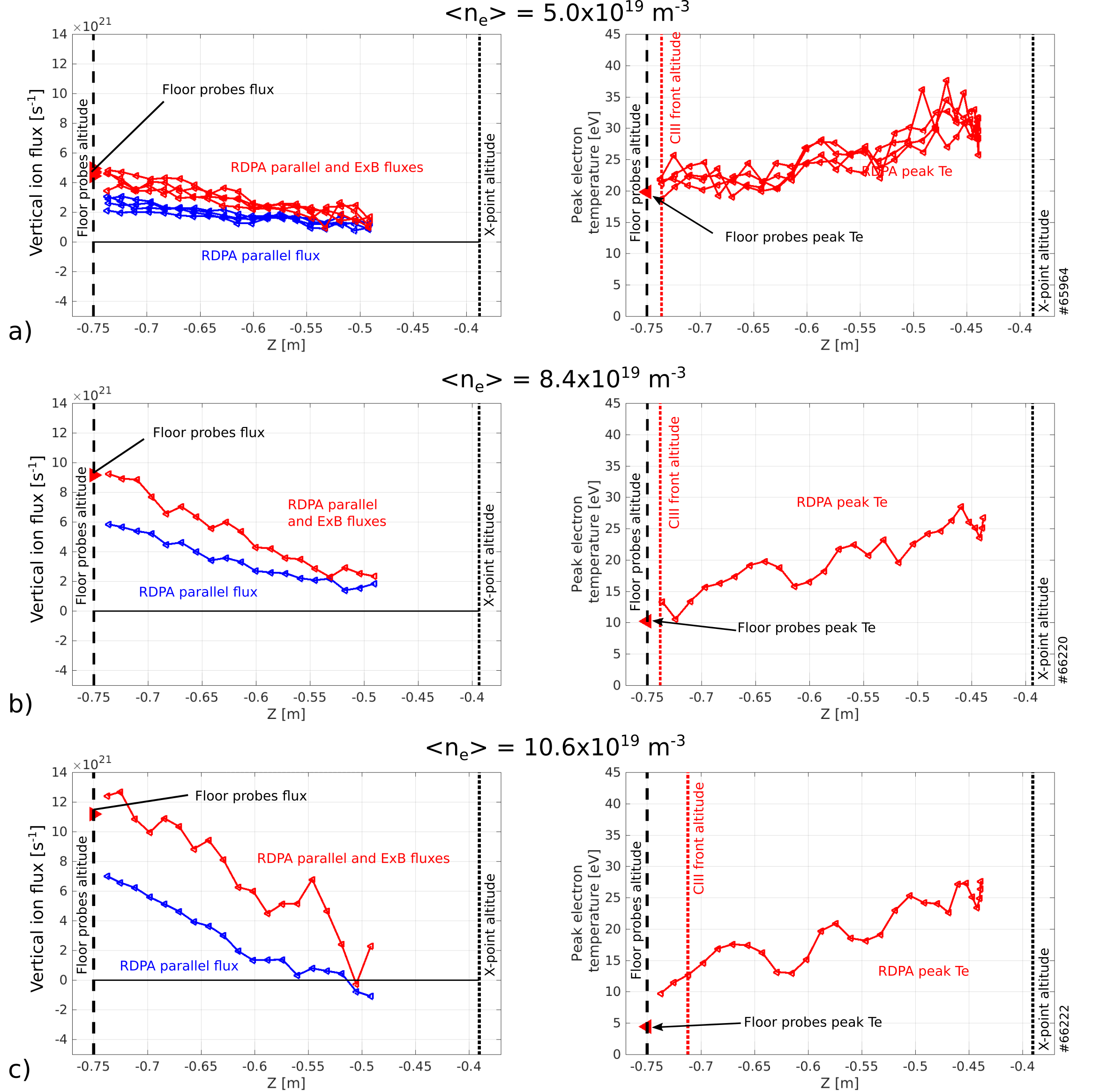}
\caption{\label{fig:figure_core_density_1_merged} 1D divertor particle balance (left column) and peak $T_e$ (right column) along the outer divertor leg for different core densities: {a)} Low core density reference \#65964 $\langle n_e \rangle \approx 5.0 \times 10^{19}~\mathrm{m^{-3}}$, {b)} Intermediate core density reference \#66220 $\langle n_e \rangle\approx 8.4 \times 10^{19}~\mathrm{m^{-3}}$ and {c)} High core density reference \#66222 $\langle n_e \rangle\approx 10.6 \times 10^{19}~\mathrm{m^{-3}}$. Blue curves in the left column show total vertical ion fluxes to the target due to parallel flows only, while the red curves include the contribution from the $E\times B$ velocity. All discharges without baffles.}
\end{figure*}

We perform a 1D particle balance analysis along the divertor leg, obtained by integrating the vertical ion flux densities due to parallel and $E\times B$ flows, obtained from the RDPA. The integration is performed both along the radial and the toroidal directions. Thus, the result does not depend on effects due to radial cross-field transport, as radial transport spreads the profile but does not change the total poloidal ion flux along the leg. Only source and sink terms, such as volumetric ionization and recombination, can change it. Figure \ref{fig:integration_drawing} shows an illustration of the integration, performed in the relevant geometry.

Figure $\ref{fig:figure_core_density_1_merged}$ shows data for three non baffled experiments with different core densities. The data reproducibility between separate RDPA plunges is generally good, as shown in Figure \ref{fig:figure_core_density_1_merged}{a)}, where $4$ curves from two up- and two down-ward plunges are superposed.

\begin{figure}
\centering
\includegraphics[width=\linewidth]{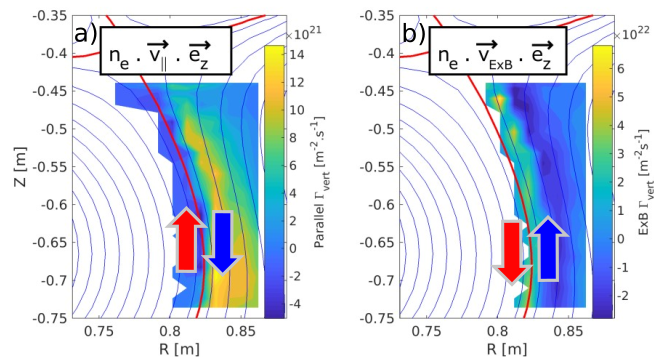}
\caption{\label{fig:figure_reversal} RDPA measurements for the low density reference without baffles (shot 65964 in Figure \ref{fig:figure_core_density_1_merged}a) of: {a)} Vertical particle flux density from the parallel Mach number measurement and {b)} Vertical $E\times B$ particle flux density.}
\end{figure}

The low density reference (Figure \ref{fig:figure_core_density_1_merged}{a)}) has a non-negligible ion flux from the upstream SOL (above the RDPA coverage), $\iint \Gamma_{upstream}dS / \iint \Gamma_{target}dS\approx 20 \%$. As the RDPA does not reach to the X-point, it is possible that the real contribution from the upstream SOL is close to zero or even negative (net flow of particles towards the inner target through the common and/or the private flux regions). The plasma shows a high peak electron temperature $T_{e,max}\approx 30~\mathrm{eV}$ and a small electron temperature gradient along the divertor leg (Figure \ref{fig:figure_core_density_1_merged}{a)}, right panel), as expected from efficient electron heat conductivity for this high electron temperature. This high electron temperature and a relatively narrow radial profile width lead to an interesting consequence: the $E\times B$ flow contribution to the poloidal particle flux is greater than the one due to the parallel flow. As one could expect from Bohm-Chodura sheath boundary conditions including drifts \cite{Stangeby2000}, the parallel flow even reverses in the downward $E\times B$ region, pointing away from the target along the divertor leg, near the separatrix, as shown in Figure \ref{fig:figure_reversal}{a)}.

In all non-baffled shots, the electron temperature ($T_{e,max} \geq 10~\mathrm{eV}$), e.g. Figure \ref{fig:figure_core_density_1_merged}, is high enough to ionize neutral particles along the entire divertor leg. This is consistent with the CIII front position, that barely leaves the floor, as shown in the right column of Figure \ref{fig:figure_core_density_1_merged}. The intermediate core density and the high density references, see Figure \ref{fig:figure_core_density_1_merged}{b)} and {c)}, have higher divertor ionization than the low density case, attested by the steeper increase in the total ion flux towards the floor and the higher flux measured by the floor probes. The parallel ion flux even appears to slightly reverse near the top of the plunge at the highest density and remains close to zero when including the $E\times B$ drift contribution (Figure \ref{fig:figure_core_density_1_merged}c). Overall, the small particle fluxes entering the divertor from upstream agrees with a ``closed box'' divertor approximation \cite{Krasheninnikov1987,Krasheninnikov2017}, where most of the ionization is taken to occur in the divertor. More studies are required in this regard on TCV, taking advantage of upstream fast scanning probe measurements to determine the particle source originating from the core and the upstream SOL.

\subsection{The influence of core density on the divertor particle balance - with baffles}\label{sec:particle_balance_baffles}

\begin{figure*}
\centering
\includegraphics[width=\linewidth]{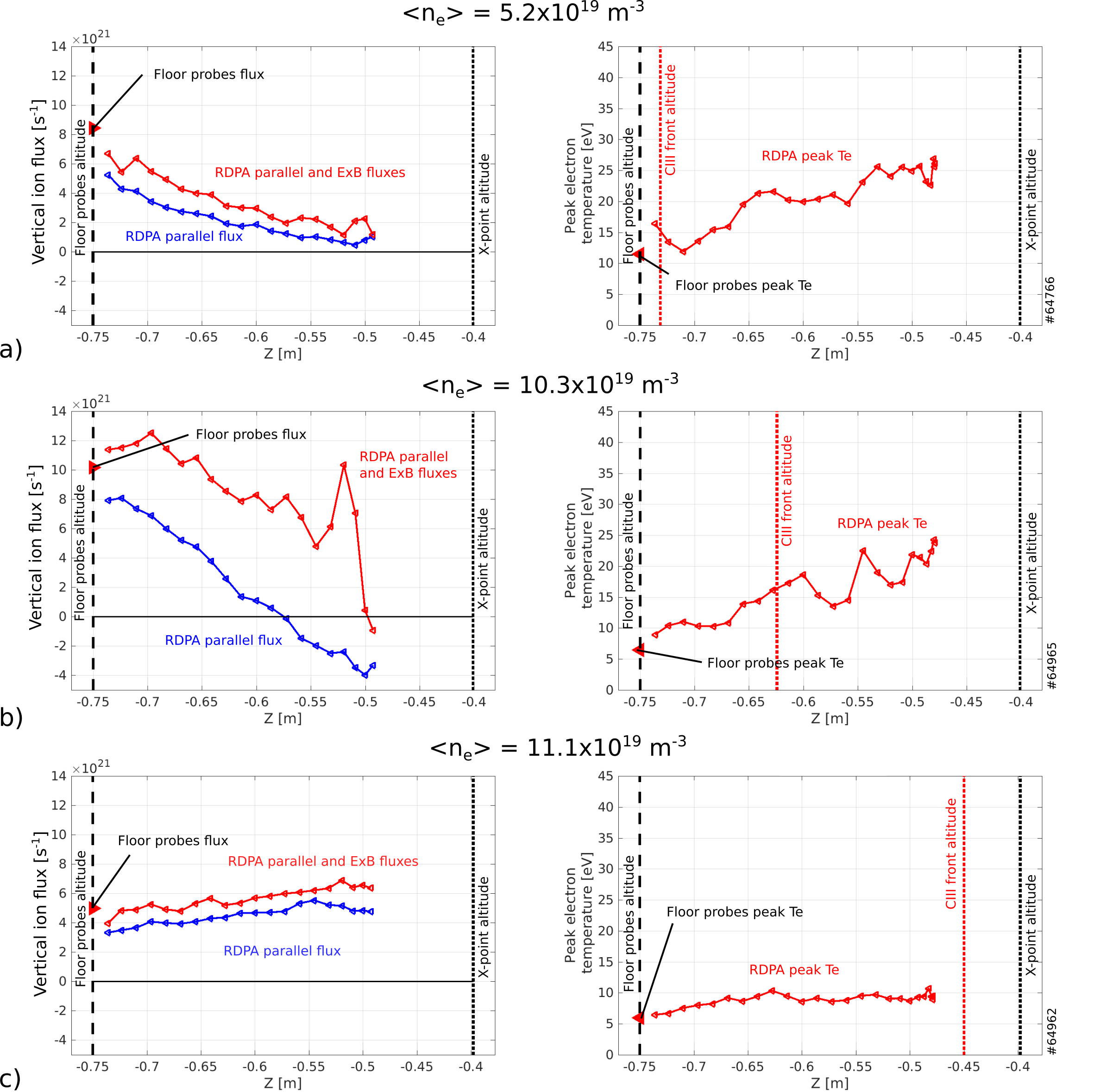}
\caption{\label{fig:figure_core_density_fluct_2_merged} Equivalent to Figure \ref{fig:figure_core_density_1_merged} for baffled discharges: {a)} Low core density references \#64766 $\langle n_e \rangle\approx 5.2 \times 10^{19}~\mathrm{m^{-3}}$, {b)} Intermediate core density reference \#64965 $\langle n_e \rangle\approx 10.3 \times 10^{19}~\mathrm{m^{-3}}$ and {c)} High core density reference \#64962 $\langle n_e \rangle\approx 11.1 \times 10^{19}~\mathrm{m^{-3}}$ (beyond roll-over). Blue curves in the left column show total vertical ion fluxes to the target due to parallel flows only, while the red curves include the contribution from the $E\times B$ velocity.}
\end{figure*}

\begin{figure*}
\centering
\includegraphics[width=\linewidth]{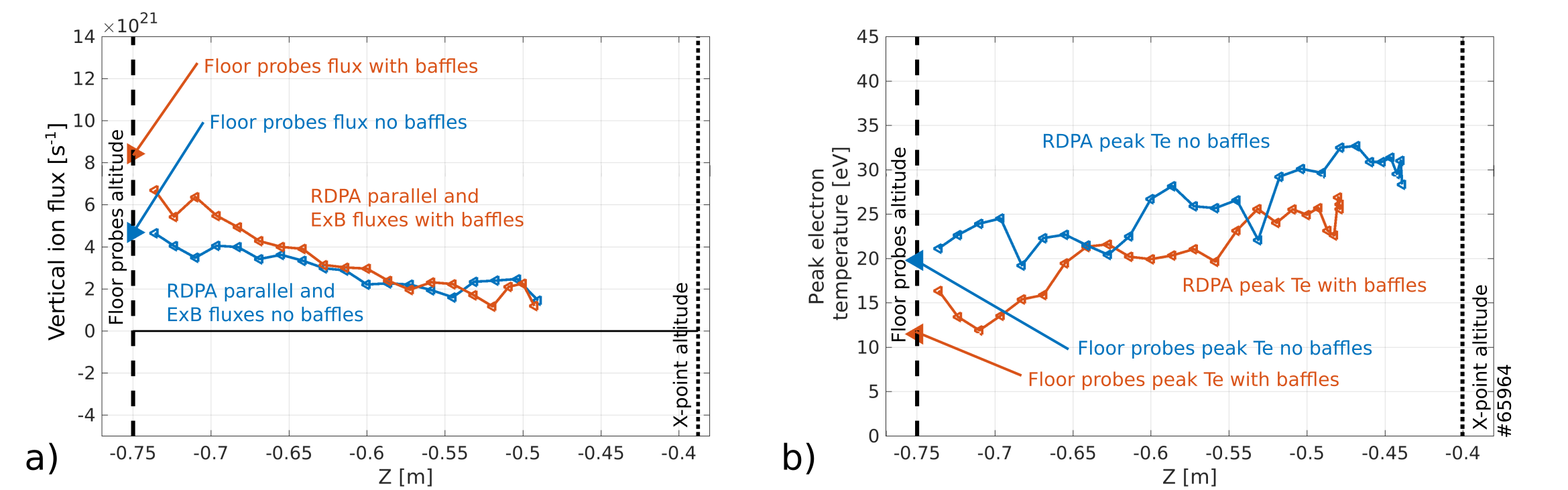}
\caption{\label{fig:figure_core_density_21_2020} Low core density discharges without baffles \#65964 ($\langle n_e \rangle\approx 5.0 \times 10^{19}~\mathrm{m^{-3}}$) and with baffles \#64766 ($\langle n_e \rangle\approx 5.2 \times 10^{19}~\mathrm{m^{-3}}$). {a)} Total ion flux to the outer target including both parallel flow and $E\times B$ contributions and {b)} Peak electron temperature value.}
\end{figure*}

Figure \ref{fig:figure_core_density_fluct_2_merged} shows RDPA results (integrated ion flux and peak electron temperature) for three baffled experiments with increasing core density. The fraction of the target ion flux from the upstream SOL (above the RDPA coverage) is $\iint \Gamma_{upstream}dS / \iint \Gamma_{target}dS\approx 15\%$ for the low-density reference {a)}. There is a high peak electron temperature at the top of the plunge and a substantial temperature gradient from $T_{e,max} \approx 25~\mathrm{eV}$ to $T_{e,max} \approx 12~\mathrm{eV}$. The CIII front position remains close to the target, similarly to the non-baffled discharges.

The intermediate core density reference, Figure $\ref{fig:figure_core_density_fluct_2_merged}${b)}, has the highest divertor ionization and no significant ion flux from the upstream SOL. The parallel flux is directed upwards (negative parallel flux) at the top of the RDPA plunge and the total flux (including the $E\times B$ drift) is close to zero. We note that the parallel flux is dominantly reversed in the region of downward $E\times B$ velocity (not shown), which is close to the separatrix in these discharges. The downward $E\times B$ transport plays a major role in the upper region of the divertor leg, where the temperature radial gradient is significantly higher than at the target. The electron temperature is marginal to neutral ionization near the target. Consequently, the total ion flux profile remains flat close to the target in the absence of a divergence term in the region. This region is below the CIII front, which is located $\approx 13~\mathrm{cm}$ above the target in this case.

The high core density reference, Figure $\ref{fig:figure_core_density_fluct_2_merged}${c)}, has a negligible divertor ionization with the entire ion flux originating from the upstream SOL, above the RDPA coverage. This is apparent from the nearly flat total vertical ion flux profile along the entire outer divertor leg. The Mach number, not shown here, is fairly constant across the entire profile and stays within $0.4$ to $0.6$. The $E \times B$ transport contribution is smaller than the parallel transport throughout the entire RDPA scan. In contrast to the other discharges, the ion flux even slightly decreases towards the target, by $\approx 20\%$ indicative of volumetric recombination. Indeed, both spectroscopic studies $\cite{Verhaegh2019DSS}$ and SOLPS simulations $\cite{Wensing2019}$ showed the presence of modest levels of recombination for detached L-mode TCV plasmas. The recombination probability exceeds that for ionization only below $T_e \approx 1.5~\mathrm{eV}$. However, molecular activated recombination (MAR) has been suggested as a possible candidate to explain the missing ion flux \cite{Kukushkin2017,Hollmann2006} at higher electron temperatures, up to $T_e \approx 3~\mathrm{eV}$. Langmuir probe floor measurements indicate $T_e\approx 6~\mathrm{eV}$ here. There is, however, the possibility of over-estimation of $T_e$ by the LPs in detached conditions \cite{Fevrier2018}, so the real value could be considerably lower. The CIII front is $\approx 5~\mathrm{cm}$ below the X-point, beyond the RDPA coverage in these discharges.

\subsection{The effect of baffles for a fixed core density}
\label{sec:particle_balance_baffles_and_no_baffle}

The effect of the baffles on detachment onset, already shown in Figure \ref{fig:density_comparison}, were clearly mirrored in the particle balance shown in Figures \ref{fig:figure_core_density_1_merged} and \ref{fig:figure_core_density_fluct_2_merged}. Here, we contrast the lowest density discharges, for which a good match of the core quantities was achieved, see Figure \ref{fig:figure_time_traces_RDPA_effect}. Clear differences between the non-baffled and baffled discharges appear only in the radiated power, the gas flux required for fueling and the divertor neutral pressure. Bolometry indicates that the increase in radiated power mainly originates from the divertor (inner leg, outer leg and X-point region), whose lower temperatures and higher densities with baffles can be expected to enhance both the carbon impurity and hydrogenic radiated power. The carbon concentration for both discharges is, however, unknown and may also be a key parameter in explaining the radiated power difference.

This comparison clearly highlights the higher divertor ionization of the baffled case, Figure \ref{fig:figure_core_density_21_2020}{a)}: more flux arrives at the target. Similar, or even slightly smaller, fluxes arrive from upstream. Such higher ionization level is expected \cite{Wensing2019} as the higher neutral density combined with sufficient plasma temperatures lead to stronger ionization. Lower electron temperatures with baffles is also apparent in Figure \ref{fig:figure_core_density_21_2020}{b)}.

\section{Fluctuation results} \label{sec:fluctuation_results}
\begin{figure*}
\centering
\includegraphics[width=\linewidth]{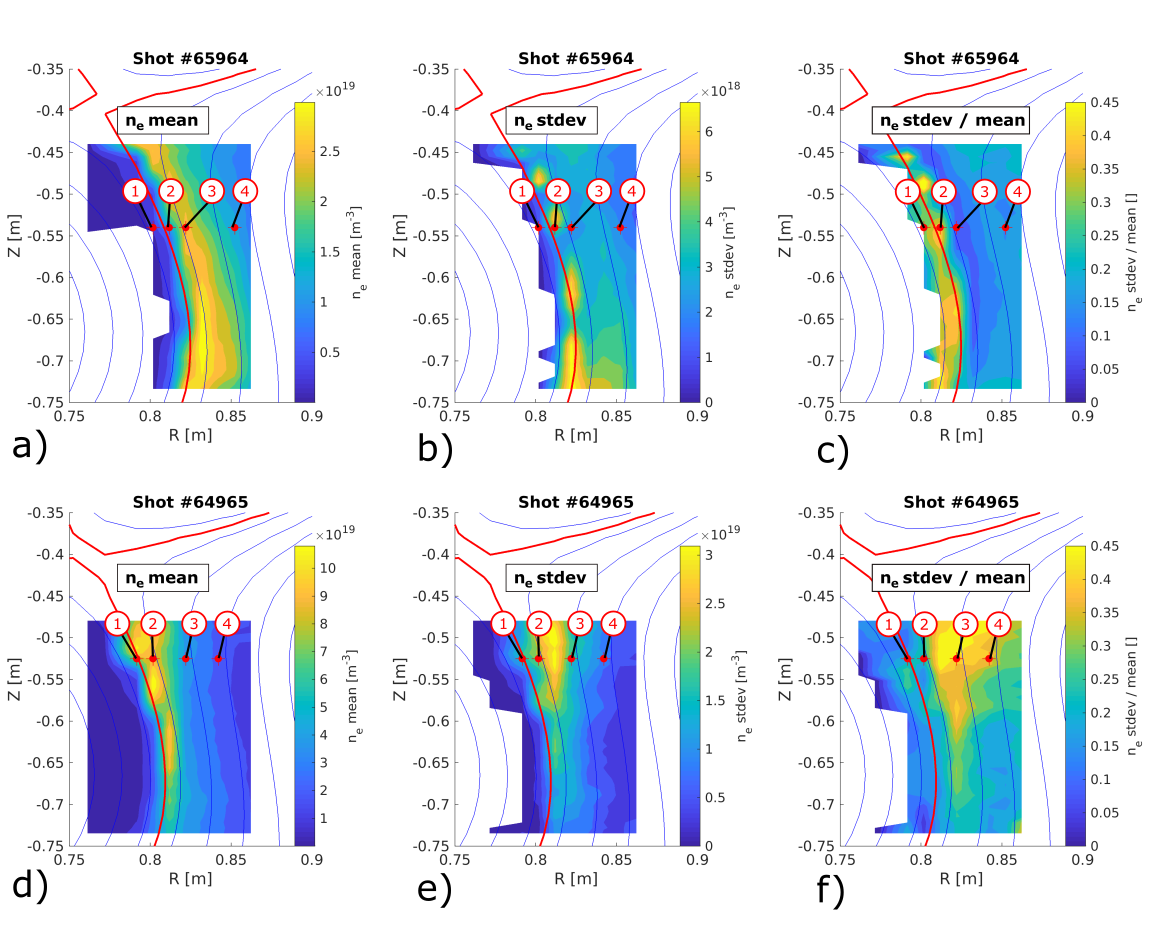}
\caption{Data from the low density non baffled shot \#65964 and the high density baffled shot \#64965 to illustrate the density fluctuation profiles in various condition: {a)} and {d)} time averaged electron density, {b)} and {e):} standard deviation of the density and {c)} and {f):} relative density fluctuation level (standard deviation normalized by the time averaged density).}
\label{fig:reference_density_fluctuation}
\end{figure*}

We have shown how poloidal particle transport in the divertor contains strong contributions from flows along the magnetic field and perpendicular to it, due to steady state $E\times B$ drifts. A full description of the divertor processes must, however, also include turbulence dynamics. Ongoing studies combine numerical simulations \cite{Paruta2018,Giacomin2020,OliveiraDiego2021,Wuthrich2021} and dedicated experiments, including RDPA measurements.

While these studies will generate further exploration, the purpose of this section is to provide a description of the fluctuation levels in the discharges described in the previous sections. Discharges with a wide range of divertor regimes are studied here: a strongly attached, low density, non-baffled discharge (\#65964 in Figures \ref{fig:reference_density_fluctuation} {a)}, {b)} and {c)}), and a high density, baffled, discharge (\#64965 in Figures \ref{fig:reference_density_fluctuation}{d)}, {e)} and {f)}), which is at the onset of detachment (note that these are the same discharges presented in Figure \ref{fig:figure_core_density_1_merged}{a)} and Figure \ref{fig:figure_core_density_21_2020}{b)}).

\begin{figure*}
\centering
\includegraphics[width=\linewidth]{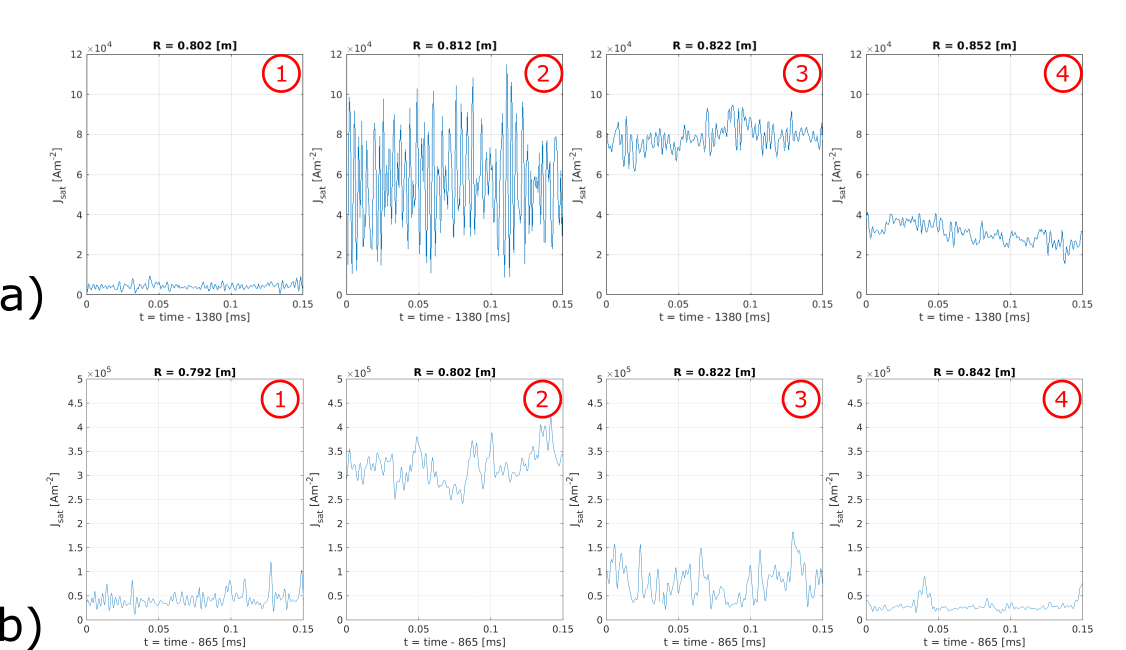}
\caption{
{a)} Ion saturation current density time traces from the low density shot without baffles \#65964: signal from the upstream probe tips of the RDPA Mach probes (the labels $\raisebox{.5pt}{\textcircled{\raisebox{-.9pt} {1}}}$ to $\raisebox{.5pt}{\textcircled{\raisebox{-.9pt} {4}}}$ and the positions of the corresponding data point in the poloidal plane are shown in Figure \ref{fig:reference_density_fluctuation}{a)}).
{b)} Ion saturation current density time traces from the high density shot with baffles \#64965: signal from the upstream probe tips of the RDPA Mach probes (the labels $\raisebox{.5pt}{\textcircled{\raisebox{-.9pt} {1}}}$ to $\raisebox{.5pt}{\textcircled{\raisebox{-.9pt} {4}}}$ and the positions of the corresponding data point in the poloidal plane are shown in Figure \ref{fig:reference_density_fluctuation}{d)}).
}
\label{fig:density_fluctuation_raw_signal}
\end{figure*}

In the low density example, relative density fluctuation levels are almost constant along a flux surface in the divertor leg, Figure \ref{fig:reference_density_fluctuation}{c)}. The data reveals four distinct regions to facilitate the description, Figures \ref{fig:density_fluctuation_raw_signal}{a)} and \ref{fig:reference_density_fluctuation}{c)}:
$\raisebox{.5pt}{\textcircled{\raisebox{-.9pt} {1}}}$ The private flux region, characterized by a relatively quiescent and faint plasma. 
$\raisebox{.5pt}{\textcircled{\raisebox{-.9pt} {2}}}$ The proximity of the separatrix, on the HFS of the electron temperature peak (region of strong downward $E\times B$ drift in these reversed field discharges), where a high fluctuation level is, surprisingly, observed, with $\widetilde{n_e}/\langle n_e \rangle \approx 0.5$. The origin of this high fluctuation intensity is unknown. $\raisebox{.5pt}{\textcircled{\raisebox{-.9pt} {3}}}$ Near the electron temperature peak. The density fluctuation level in this region is the lowest $\widetilde{n_e}/\langle n_e \rangle \approx 0.1$. 
$\raisebox{.5pt}{\textcircled{\raisebox{-.9pt} {4}}}$ To the LFS of the electron temperature peak. The density fluctuation level increases gradually from $\widetilde{n_e}/\langle n_e \rangle \approx 0.1$ to $\widetilde{n_e}/\langle n_e \rangle \approx 0.25$ in the far SOL. The higher fluctuation level in the far SOL, compared to region $\raisebox{.5pt}{\textcircled{\raisebox{-.9pt} {3}}}$, is believed to result from reconnection with upstream turbulence. This reconnection process was recently shown in the far SOL of TCV in similar, low collisionnality, discharges by simultaneously tracking convective cells with the Gas Puff Imaging (GPI) diagnostic \cite{Offeddu2019} and RDPA mapped along field lines.

In the high-density example, the turbulence behavior differs strongly from the low collisionnality case. The relative fluctuation levels change along flux surfaces, as shown in Figure $\ref{fig:reference_density_fluctuation}${f)}. The data is, once again, divided in four regions, at different positions, to facilitate the description, Figure \ref{fig:density_fluctuation_raw_signal}{b)} and Figure \ref{fig:reference_density_fluctuation}{f)}:
$\raisebox{.5pt}{\textcircled{\raisebox{-.9pt} {1}}}$ As in the low density reference, the private plasma is characterized by a relatively quiescent region with very little plasma. $\raisebox{.5pt}{\textcircled{\raisebox{-.9pt} {2}}}$ Close to the separatrix, on the HFS of the electron temperature peak, low fluctuation levels, compared to the low density reference, of $\widetilde{n_e}/\langle n_e \rangle \approx 0.2$, are observed. The narrow, high fluctuation level feature to the HFS of the $T_e$ peak, seen in the low-density reference, is thus absent with increasing collisionnality. $\raisebox{.5pt}{\textcircled{\raisebox{-.9pt} {3}}}$ The low fluctuation region near the maximum of the $T_e$ profile present in the low-density case is absent for these higher density conditions Indeed, in the near SOL region, close to the X-point height, where fluctuations are believed to be related to the upstream turbulence, fluctuation levels of $\widetilde{n_e}/\langle n_e \rangle \approx 0.45$ are observed. This value drops to $\widetilde{n_e}/\langle n_e \rangle \approx 0.2$ near the target. This drop coincides with the electron temperature drop shown in Figure \ref{fig:figure_core_density_fluct_2_merged}{b)} and the associated increasing collisionality. $\raisebox{.5pt}{\textcircled{\raisebox{-.9pt} {4}}}$ The far-SOL, where turbulent events are much weaker.

These measurements thus clearly show how the divertor fluctuation behavior strongly varies between attached and detached divertor conditions, both radially and poloidally. The consequences of theses changes on divertor turbulence transport and target profile broadening is an important subject for future studies. 


\section{Conclusion} \label{sec:conclusion}
This study reports detailed 2D Langmuir probe measurements across a large part of the divertor region in TCV. The measured quantities include plasma density, temperature, potential and parallel Mach number. These are used to estimate vertical particle flux densities associated both with parallel flows and $E \times B$ flows to generate a particle balance in the divertor. This method was applied to Ohmic L-mode plasmas in both baffled and non-baffled divertor configurations. A range of divertor regimes were accessed by varying the plasma line-averaged density. To access high-density conditions, all experiments were performed at a relatively high plasma current of $\approx 320~\mathrm{kA}$ and with an unfavorable ion-grad B drift direction to avoid H-mode transitions.

Baffles are found to substantially increase the divertor neutral pressure and facilitate the detachment onset. These effects were similar but stronger than recent experiments with a lower plasma current\cite{Reimerdes2021,Fevrier2021}. Results from density ramps and constant density discharges agree well, at corresponding densities, in terms of the integrated ion flux, the CIII front position and the divertor neutral pressure, allowing some reduction in the required number of experimental discharges.

This study reveals that the contributions from the $E \times B$ drift to the poloidal ion flux can be comparable, and sometimes larger, than that from the ion flux along the magnetic field lines, consistent with reported observations from DIII-D \cite{Boedo2000,Boedo1999,Jarvinen2019}. A detailed particle balance in the divertor reveals that, in most cases, and with and without baffles, most of the particle flux to the outer target results from ionization along the outer divertor leg. This supports the closed-box approximation frequently used in detachment models \cite{Krasheninnikov1987,Krasheninnikov2017} and previously inferred on TCV from spectroscopic measurements \cite{Verhaegh2019}. Close to detachment onset, in these reversed field conditions, the integrated parallel particle flux can even ``reverse'' just below the X-point, i.e. ions flow upstream along the magnetic field, while the total flow, including the $E\times B$ drift, remains positive, i.e. directed towards the target. In the most detached conditions, the divertor leg becomes too cold for any significant ionization to occur. In this extreme case, achieved only with baffles in these experiments, the entire particle source is located in the proximity of, or even above, the X-point, and the flux along the divertor leg can even slightly decrease as it approaches the target, ascribed to plasma recombination. A close match between the attached and lowest density discharges, with and without baffles, was achieved in terms of upstream properties such as line-averaged density, that revealed a substantial increase in the divertor ionization level with baffles. Additionally, divertor electron temperatures were reduced by $\approx 35\%$.

Observations regarding divertor density fluctuation measurements in these plasmas include the development of a poloidal gradient in the fluctuation levels, with increasing collisionnality, and the surprising presence of a narrow, high fluctuation level region in the downward $E\times B$ region for strongly attached, reversed field discharges. Many future research avenues are now available to further improve the description of the divertor fluctuation properties using the RDPA such as: cross correlation with $J_{sat}$ and $V_{float}$ signals using the split Mach tip configuration, the assessment of the heat and particle profile spreading along the divertor leg, or the study of the divertor fluctuation and time-averaged profile properties in advanced divertor geometries. 

\section{Acknowledgments} \label{sec:acknowledgments}
This work was supported in part by the Swiss National Science Foundation. This work has been carried out within the framework of the EUROfusion Consortium, funded by the European Union via the Euratom Research and Training Programme (Grant Agreement No 101052200 — EUROfusion). Views and opinions expressed are however those of the author(s) only and do not necessarily reflect those of the European Union or the European Commission. Neither the European Union nor the European Commission can be held responsible for them. This work was supported in part by the US Department of Energy under Award Number DE-SC0010529.


\section{References}

\begin{thebibliography}{10}

\bibitem{Reimerdes_2022}
H.~Reimerdes, M.~Agostini, E.~Alessi, S.~Alberti, Y.~Andrebe, H.~Arnichand,
  J.~Balbin, F.~Bagnato, M.~Baquero-Ruiz, M.~Bernert, W.~Bin, P.~Blanchard,
  T.C. Blanken, J.A. Boedo, D.~Brida, S.~Brunner, C.~Bogar, O.~Bogar,
  T.~Bolzonella, F.~Bombarda, F.~Bouquey, C.~Bowman, D.~Brunetti, J.~Buermans,
  H.~Bufferand, L.~Calacci, Y.~Camenen, S.~Carli, D.~Carnevale, F.~Carpanese,
  F.~Causa, J.~Cavalier, M.~Cavedon, J.A. Cazabonne, J.~Cerovsky, R.~Chandra,
  A.~Chandrarajan Jayalekshmi, O.~Chellaï, P.~Chmielewski, D.~Choi,
  G.~Ciraolo, I.G.J. Classen, S.~Coda, C.~Colandrea, A.~Dal Molin, P.~David,
  M.R. de~Baar, J.~Decker, W.~Dekeyser, H.~de~Oliveira, D.~Douai, M.~Dreval,
  M.G. Dunne, B.P. Duval, S.~Elmore, O.~Embreus, F.~Eriksson, M.~Faitsch,
  G.~Falchetto, M.~Farnik, A.~Fasoli, N.~Fedorczak, F.~Felici,
  O.~F{\'{e}}vrier, O.~Ficker, A.~Fil, M.~Fontana, E.~Fransson, L.~Frassinetti,
  I.~Furno, D.S. Gahle, D.~Galassi, K.~Galazka, C.~Galperti, S.~Garavaglia,
  M.~Garcia-Munoz, B.~Geiger, M.~Giacomin, G.~Giruzzi, M.~Gobbin,
  T.~Golfinopoulos, T.~Goodman, S.~Gorno, G.~Granucci, J.P. Graves, M.~Griener,
  M.~Gruca, T.~Gyergyek, R.~Haelterman, A.~Hakola, W.~Han, T.~Happel,
  G.~Harrer, J.R. Harrison, S.~Henderson, G.M.D. Hogeweij, J.-P. Hogge,
  M.~Hoppe, J.~Horacek, Z.~Huang, A.~Iantchenko, P.~Innocente, K.~Insulander
  Björk, C.~Ionita-Schrittweiser, H.~Isliker, A.~Jardin, R.J.E. Jaspers,
  R.~Karimov, A.N. Karpushov, Y.~Kazakov, M.~Komm, M.~Kong, J.~Kovacic,
  O.~Krutkin, O.~Kudlacek, U.~Kumar, R.~Kwiatkowski, B.~Labit, L.~Laguardia,
  J.T. Lammers, E.~Laribi, E.~Laszynska, A.~Lazaros, O.~Linder, B.~Linehan,
  B.~Lipschultz, X.~Llobet, J.~Loizu, T.~Lunt, E.~Macusova, Y.~Marandet,
  M.~Maraschek, G.~Marceca, C.~Marchetto, S.~Marchioni, E.S. Marmar, Y.~Martin,
  L.~Martinelli, F.~Matos, R.~Maurizio, M.-L. Mayoral, D.~Mazon, V.~Menkovski,
  A.~Merle, G.~Merlo, H.~Meyer, K.~Mikszuta-Michalik, P.A.~Molina Cabrera,
  J.~Morales, J.-M. Moret, A.~Moro, D.~Moulton, H.~Muhammed, O.~Myatra,
  D.~Mykytchuk, F.~Napoli, R.D. Nem, A.H. Nielsen, M.~Nocente, S.~Nowak,
  N.~Offeddu, J.~Olsen, F.P. Orsitto, O.~Pan, G.~Papp, A.~Pau, A.~Perek,
  F.~Pesamosca, Y.~Peysson, L.~Pigatto, C.~Piron, M.~Poradzinski, L.~Porte,
  T.~Pütterich, M.~Rabinski, H.~Raj, J.J. Rasmussen, G.A. Ratt{\'{a}},
  T.~Ravensbergen, D.~Ricci, P.~Ricci, N.~Rispoli, F.~Riva, J.F.
  Rivero-Rodriguez, M.~Salewski, O.~Sauter, B.S. Schmidt, R.~Schrittweiser,
  S.~Sharapov, U.A. Sheikh, B.~Sieglin, M.~Silva, A.~Smolders, A.~Snicker,
  C.~Sozzi, M.~Spolaore, A.~Stagni, L.~Stipani, G.~Sun, T.~Tala, P.~Tamain,
  K.~Tanaka, A.~Tema Biwole, D.~Terranova, J.L. Terry, D.~Testa, C.~Theiler,
  A.~Thornton, A.~Thrys{\o}e, H.~Torreblanca, C.K. Tsui, D.~Vaccaro, M.~Vallar,
  M.~van Berkel, D.~Van Eester, R.J.R. van Kampen, S.~Van Mulders, K.~Verhaegh,
  T.~Verhaeghe, N.~Vianello, F.~Villone, E.~Viezzer, B.~Vincent,
  I.~Voitsekhovitch, N.M.T. Vu, N.~Walkden, T.~Wauters, H.~Weisen, N.~Wendler,
  M.~Wensing, F.~Widmer, S.~Wiesen, M.~Wischmeier, T.A. Wijkamp,
  D.~Wünderlich, C.~Wüthrich, V.~Yanovskiy, J.~Zebrowski, and the EUROfusion
  MST1~Team.
\newblock Overview of the {TCV} tokamak experimental programme.
\newblock {\em Nuclear Fusion}, 62(4):042018, mar 2022.

\bibitem{Perek2019}
A.~{Perek}, W.~A.~J. {Vijvers}, Y.~{Andrebe}, I.~G.~J. {Classen}, B.~P.
  {Duval}, C.~{Galperti}, J.~R. {Harrison}, B.~L. {Linehan}, T.~{Ravensbergen},
  K.~{Verhaegh}, M.~R. {de Baar}, and EUROfusion MST1~Team {TCV Team}.
\newblock {MANTIS: A real-time quantitative multispectral imaging system for
  fusion plasmas}.
\newblock {\em Review of Scientific Instruments}, 90(12):123514, December 2019.

\bibitem{DeOliveiraThesis}
H.~De~Oliveira.
\newblock {\em A fast-moving Langmuir probe array for the divertor of the
  {Tokamak à Configuration Variable}}.
\newblock PhD thesis, EPFL, SB, Lausanne, 2021.

\bibitem{DeOliveira2021}
H.~{De Oliveira}, C.~{Theiler}, H.~{Elaian}, and {TCV Team}.
\newblock {A fast-reciprocating probe array for two-dimensional measurements in
  the divertor region of the Tokamak {\`a} configuration variable}.
\newblock {\em Review of Scientific Instruments}, 92(4):043547, April 2021.

\bibitem{Reimerdes2017}
H.~Reimerdes, S.~Alberti, P.~Blanchard, P.~Bruzzone, R.~Chavan, S.~Coda, B.~P.
  Duval, A.~Fasoli, B.~Labit, B.~Lipschultz, T.~Lunt, Y.~Martin, J.~M. Moret,
  U.~Sheikh, B.~Sudki, D.~Testa, C.~Theiler, M.~Toussaint, D.~Uglietti,
  N.~Vianello, and M.~Wischmeier.
\newblock {TCV} divertor upgrade for alternative magnetic configurations.
\newblock {\em Nuclear Materials And Energy}, 12:6. 1106--1111, 2017.

\bibitem{Fasoli2020}
A.~{Fasoli}, H.~{Reimerdes}, S.~{Alberti}, M.~{Baquero-Ruiz}, B.~P. {Duval},
  E.~{Havlikova}, A.~{Karpushov}, J.~M. {Moret}, M.~{Toussaint}, H.~{Elaian},
  M.~{Silva}, C.~{Theiler}, D.~{Vaccaro}, and {the TCV team}.
\newblock {TCV heating and divertor upgrades}.
\newblock {\em Nuclear Fusion}, 60(1):016019, January 2020.

\bibitem{Boedo2000}
J.~A. {Boedo}, M.~J. {Schaffer}, R.~{Maingi}, and C.~J. {Lasnier}.
\newblock {Electric field-induced plasma convection in tokamak divertors}.
\newblock {\em Physics of Plasmas}, 7(4):1075--1078, April 2000.

\bibitem{Boedo1999}
J.~A. {Boedo}, R.~{Lehmer}, R.~A. {Moyer}, J.~G. {Watkins}, G.~D. {Porter},
  T.~E. {Evans}, A.~W. {Leonard}, and M.~J. {Schaffer}.
\newblock {Measurements of flows in the DIII-D divertor by Mach probes}.
\newblock {\em Journal of Nuclear Materials}, 266:783--787, January 1999.

\bibitem{Jarvinen2019}
A.E. Jaervinen, S.L. Allen, A.W. Leonard, A.G. McLean, A.L. Moser, T.D.
  Rognlien, and C.M. Samuell.
\newblock Role of poloidal exb drift in divertor heat transport in diii-d.
\newblock {\em Contributions to Plasma Physics}, 60(5-6):e201900111, 2020.
\newblock e201900111 ctpp.201900111.

\bibitem{Krasheninnikov1987}
S.I. Krasheninnikov, A.S. Kukushkin, V.I. Pistunovich, and V.A. Pozharov.
\newblock Self-sustained oscillations in the divertor plasma.
\newblock {\em Nuclear Fusion}, 27(11):1805--1816, November 1987.

\bibitem{Krasheninnikov2017}
S.~I. {Krasheninnikov}, A.~S. {Kukushkin}, Wonjae {Lee}, A.~A. {Phsenov}, R.~D.
  {Smirnov}, A.~I. {Smolyakov}, A.~A. {Stepanenko}, and Yanzeng {Zhang}.
\newblock {Edge and divertor plasma: detachment, stability, and plasma-wall
  interactions}.
\newblock {\em Nuclear Fusion}, 57(10):102010, October 2017.

\bibitem{Theiler2017}
C.~{Theiler}, B.~{Lipschultz}, J.~{Harrison}, B.~{Labit}, H.~{Reimerdes},
  C.~{Tsui}, W.~A.~J. {Vijvers}, J.~A. {Boedo}, B.~P. {Duval}, S.~{Elmore},
  P.~{Innocente}, U.~{Kruezi}, T.~{Lunt}, R.~{Maurizio}, F.~{Nespoli},
  U.~{Sheikh}, A.~J. {Thornton}, S.~H.~M. {van Limpt}, K.~{Verhaegh},
  N.~{Vianello}, {the TCV Team}, and {the EUROfusion MST1 Team}.
\newblock {Results from recent detachment experiments in alternative divertor
  configurations on TCV}.
\newblock {\em Nuclear Fusion}, 57(7):072008, July 2017.

\bibitem{Reimerdes2021}
H.~Reimerdes, B.P. Duval, H.~Elaian, A.~Fasoli, O.~F{\'{e}}vrier, C.~Theiler,
  F.~Bagnato, M.~Baquero-Ruiz, P.~Blanchard, D.~Brida, C.~Colandrea, H.~De
  Oliveira, D.~Galassi, S.~Gorno, S.~Henderson, M.~Komm, B.~Linehan,
  L.~Martinelli, R.~Maurizio, J.-M. Moret, A.~Perek, H.~Raj, U.~Sheikh,
  D.~Testa, M.~Toussaint, C.K. Tsui, M.~Wensing, the TCV~team, and the
  EUROfusion MST1~team.
\newblock Initial {TCV} operation with a baffled divertor.
\newblock {\em Nuclear Fusion}, 61(2):024002, January 2021.

\bibitem{Fevrier2021}
O.~Février, H.~Reimerdes, C.~Theiler, D.~Brida, C.~Colandrea, H.~{De
  Oliveira}, B.P. Duval, D.~Galassi, S.~Gorno, S.~Henderson, M.~Komm, B.~Labit,
  B.~Linehan, L.~Martinelli, A.~Perek, H.~Raj, U.~Sheikh, C.K. Tsui, and
  M.~Wensing.
\newblock Divertor closure effects on the {TCV} boundary plasma.
\newblock {\em Nuclear Materials and Energy}, 27:100977, 2021.

\bibitem{Fevrier2018}
O.~{F{\'e}vrier}, C.~{Theiler}, H.~{De Oliveira}, B.~{Labit}, N.~{Fedorczak},
  and A.~{Baillod}.
\newblock {Analysis of wall-embedded Langmuir probe signals in different
  conditions on the Tokamak {\`a} Configuration Variable}.
\newblock {\em Review of Scientific Instruments}, 89(5):053502, May 2018.

\bibitem{DeOliveira2019}
H.~De Oliveira, P.~Marmillod, C.~Theiler, R.~Chavan, O.~F{\'{e}}vrier,
  B.~Labit, P.~Lavanchy, B.~Marl{\'{e}}taz, and R.~A.~Pitts and.
\newblock Langmuir probe electronics upgrade on the tokamak {\`{a}}
  configuration variable.
\newblock {\em Review of Scientific Instruments}, 90(8):083502, August 2019.

\bibitem{Hutchinson2005}
I.~H. {Hutchinson}.
\newblock {\em {Principles of Plasma Diagnostics}}.
\newblock Cambridge University Press, 2005.

\bibitem{Stangeby2000}
P.~C. {Stangeby}.
\newblock {\em {The Plasma Boundary of Magnetic Fusion Devices}}.
\newblock Series in Plasma Physics and Fluid Dynamics. Taylor and Francis,
  2000.

\bibitem{Hutchinson2008}
I.~H. {Hutchinson}.
\newblock {Oblique ion collection in the drift approximation: How magnetized
  Mach probes really work}.
\newblock {\em Physics of Plasmas}, 15(12):123503, December 2008.

\bibitem{Jarvinen2018}
A.~E. {Jaervinen}, S.~L. {Allen}, D.~{Eldon}, M.~E. {Fenstermacher},
  M.~{Groth}, D.~N. {Hill}, A.~W. {Leonard}, A.~G. {McLean}, G.~D. {Porter},
  T.~D. {Rognlien}, C.~M. {Samuell}, and H.~Q. {Wang}.
\newblock {ExB Flux Driven Detachment Bifurcation in the DIII-D Tokamak}.
\newblock {\em Physical Review Letters}, 121(7):075001, August 2018.

\bibitem{Nold2012}
B.~{Nold}, T.~T. {Ribeiro}, M.~{Ramisch}, Z.~{Huang}, H.~W. {M{\"u}ller}, B.~D.
  {Scott}, U.~{Stroth}, and {the ASDEX Upgrade Team}.
\newblock {Influence of temperature fluctuations on plasma turbulence
  investigations with Langmuir probes}.
\newblock {\em New Journal of Physics}, 14(6):063022, June 2012.

\bibitem{Mahdizadeh2005}
N.~{Mahdizadeh}, F.~{Greiner}, M.~{Ramisch}, U.~{Stroth}, W.~{Guttenfelder},
  C.~{Lechte}, and K.~{Rahbarnia}.
\newblock {Comparison of Langmuir and emissive probes as diagnostics for
  turbulence studies in the low-temperature plasma of the torsatron TJ-K}.
\newblock {\em Plasma Physics and Controlled Fusion}, 47(4):569--579, April
  2005.

\bibitem{Verhaegh2019DSS}
K.~{Verhaegh}, B.~{Lipschultz}, B.~P. {Duval}, A.~{Fil}, M.~{Wensing},
  C.~{Bowman}, and D.~S. {Gahle}.
\newblock {Novel inferences of ionisation and recombination for particle/power
  balance during detached discharges using deuterium Balmer line spectroscopy}.
\newblock {\em Plasma Physics and Controlled Fusion}, 61(12):125018, December
  2019.

\bibitem{Wensing2019}
M.~{Wensing}, B.~P. {Duval}, O.~{F{\'e}vrier}, A.~{Fil}, D.~{Galassi},
  E.~{Havlickova}, A.~{Perek}, H.~{Reimerdes}, C.~{Theiler}, K.~{Verhaegh},
  M.~{Wischmeier}, {the EUROfusion MST1 team}, and {the TCV team}.
\newblock {SOLPS-ITER simulations of the TCV divertor upgrade}.
\newblock {\em Plasma Physics and Controlled Fusion}, 61(8):085029, August
  2019.

\bibitem{Kukushkin2017}
A.S. Kukushkin, S.I. Krasheninnikov, A.A. Pshenov, and D.~Reiter.
\newblock Role of molecular effects in divertor plasma recombination.
\newblock {\em Nuclear Materials and Energy}, 12:984 -- 988, 2017.
\newblock Proceedings of the 22nd International Conference on Plasma Surface
  Interactions 2016, 22nd PSI.

\bibitem{Hollmann2006}
E.~M. {Hollmann}, S.~{Brezinsek}, N.~H. {Brooks}, M.~{Groth}, A.~G. {McLean},
  A.~Yu {Pigarov}, and D.~L. {Rudakov}.
\newblock {Spectroscopic measurement of atomic and molecular deuterium fluxes
  in the DIII-D plasma edge}.
\newblock {\em Plasma Physics and Controlled Fusion}, 48(8):1165--1180, August
  2006.

\bibitem{Paruta2018}
Paola Paruta, P.~Ricci, F.~Riva, C.~Wersal, C.~Beadle, and B.~Frei.
\newblock Simulation of plasma turbulence in the periphery of diverted tokamak
  by using the {GBS} code.
\newblock {\em Physics of Plasmas}, 25(11):112301, November 2018.

\bibitem{Giacomin2020}
M.~Giacomin, L.N. Stenger, and P.~Ricci.
\newblock Turbulence and flows in the plasma boundary of snowflake magnetic
  configurations.
\newblock {\em Nuclear Fusion}, 60(2):024001, January 2020.

\bibitem{OliveiraDiego2021}
Diego Sales~de Oliveira, Thomas~A Body, Davide Galassi, Christian Theiler,
  Elias Laribi, Patrick Tamain, Andreas Stegmeir, Maurizio Giacomin, Wladimir
  Zholobenko, Paolo Ricci, H.~Bufferand, J.~A. Boedo, G.~Ciraolo, C.~Colandrea,
  D.~Coster, H.~de~Oliveira, G.~Fourestey, S.~Gorno, F.~Imbeaux, F.~Jenko,
  V.~Naulin, N.~Offeddu, H.~Reimerdes, E.~Serre, C.~K. Tsui, N.~Varini,
  N.~Vianello, M.~Wiesenberger, and C.~Wüthrich.
\newblock Validation of edge turbulence codes against the {TCV-X21} diverted
  {L}-mode reference case.
\newblock {\em Nuclear Fusion}, 2022.

\bibitem{Wuthrich2021}
Curdin Wüthrich, Christian Theiler, Nicola Offeddu, Davide Galassi,
  Diego~Sales de~Oliveira, Basil Duval, Olivier Février, Theodore
  Golfinopoulos, Woonghee Han, Earl Marmar, Jim Terry, and Cedric Tsui.
\newblock X-point and divertor filament dynamics from gas puff imaging on
  {TCV}.
\newblock {\em Submitted to Nuclear Fusion}, 2022.

\bibitem{Offeddu2019}
N.~Offeddu, W.~Han, C.~Theiler, T.~Golfinopoulos, C.~Galperti, B.P. Duval,
  J.~Terry, and the TCV~Team.
\newblock Plasma edge turbulence characterization using gas puff imaging on the
  {TCV} tokamak.
\newblock {\em Submitted to Nuclear Fusion}, 2022.

\bibitem{Verhaegh2019}
K.~{Verhaegh}, B.~{Lipschultz}, B.~P. {Duval}, O.~{F{\'e}vrier}, A.~{Fil},
  C.~{Theiler}, M.~{Wensing}, C.~{Bowman}, D.~S. {Gahle}, J.~R. {Harrison},
  B.~{Labit}, C.~{Marini}, R.~{Maurizio}, H.~{de Oliveira}, H.~{Reimerdes},
  U.~{Sheikh}, C.~K. {Tsui}, N.~{Vianello}, and W.~A.~J. {Vijvers}.
\newblock {An improved understanding of the roles of atomic processes and power
  balance in divertor target ion current loss during detachment}.
\newblock {\em Nuclear Fusion}, 59(12):126038, December 2019.

\end{thebibliography}
\end{document}